\documentclass[twocolumn, 10pt, final]{autart}

\usepackage{layouts}
\usepackage{graphicx}
\DeclareGraphicsExtensions{.eps}

\usepackage{amsmath,amsfonts, bm}

\usepackage{algorithmic}

\usepackage{array}
\newcolumntype{P}[1]{>{\centering\arraybackslash}p{#1}}
\newcolumntype{M}[1]{>{\centering\arraybackslash}m{#1}}

\usepackage{arydshln}

\usepackage{booktabs}

\let\originalhdash\hdashline
\renewcommand{\hdashline}[0]{\originalhdash[2pt/2pt]}

\usepackage{textcomp}

\usepackage{enumitem}

\usepackage{siunitx}

\usepackage[labelformat=simple]{subcaption}

\usepackage{tikz}
\usetikzlibrary{calc}

\usepackage{mathtools}

\usepackage{rotating}

\usepackage{layouts}

\usepackage[shortcuts]{extdash}

\usepackage{hyperref}
\begin{document}
\begin{frontmatter}
\title{Switching and Information Exchange in Compressed Estimation of Coupled High Dimensional Processes}
\author[unsw]{K.~Narula}
\ead{k.narula@unsw.edu.au}
\author[unsw]{Jose~E.~Guivant}
\ead{j.guivant@unsw.edu.au}

\address[unsw]{School of Mechanical and Manufacturing Engineering, The University of New South Wales, Sydney 2052, NSW, Australia}

\begin{keyword}
Generalised Compressed Kalman Filter, High Dimensional Estimation, Stochastic PDEs.
\end{keyword}

\begin{abstract}
Compressed Estimation approaches, such as the Generalised Compressed Kalman Filter (GCKF), reduce the computational cost and complexity of high dimensional and high frequency data assimilation problems; usually without sacrificing optimality. Configured using adequate cores, such as the Unscented Kalman Filter (UKF), the GCKF could also treat certain non-linear cases. However, the application of a compressed estimation process is limited to a class of problems which inherently allow the estimation process to be divided, at certain intervals of time, in a subset of lower dimensional problems. This limitation prohibits applying the compressing techniques for estimating densely coupled high dimensional processes. However, those limitations can be overcome by applying proper techniques. In this paper, the concept of subsystem switching, and information exchange architecture, namely  `Exploiting Local Statistical Dependency' (ELSD), has been derived and explored, for allowing compressed estimators to mimic optimal full Gaussian estimators.  The performances of the methods have been verified through its application in solving usual types of linear Stochastic Partial Differential Equations (SPDEs). The computational advantages of using the proposed techniques have also been highlighted with recommendation of its usage over the full filter when dealing with high dimensional and high frequency data assimilation.
\end{abstract}
\end{frontmatter}
\newcommand{\ta}{t_a}
\newcommand{\tb}{t_b}
\newcommand{\transpose}[1]{{#1}^T}
\newcommand{\timenota}[1]{(#1)}
\newcommand{\Xk}[1][]{\bm{X_{#1}}}
\newcommand{\Xkh}[1][]{\hat{\bm{X}}_{\bm{#1}}}
\newcommand{\Xktime}[2][]{\Xk[#1]\timenota{#2}}
\newcommand{\Xkhtime}[2][]{\hat{\bm{X}}_{\bm{#1}}\timenota{#2}}
\newcommand{\Xkaug}[1][]{\Xk[#1]^*}
\newcommand{\XkaugGen}[1][\Xk]{{#1}^*}
\newcommand{\Xkaugtime}[2][]{\Xkaug[#1]\timenota{#2}}
\newcommand{\Xkt}[1]{\transpose{\Xk[#1]}}
\newcommand{\Xkttime}[2]{\Xkt{#1}\timenota{#2}}
\newcommand{\Xku}[1]{\Xk[#1]^u}
\newcommand{\Xkutime}[2]{\Xku{#1}\timenota{#2}}
\newcommand{\Xkut}[1]{\transpose{\Xku{#1}}}
\newcommand{\Xkuttime}[2]{\Xkut{#1}\timenota{#2}}
\newcommand{\Xo}{\Xk^{o}}
\newcommand{\Xoh}{\Xkh^{o}}
\newcommand{\Xu}{\Xk^{u}}
\newcommand{\Xuh}{\Xkh^{u}}
\newcommand{\Xoaux}{\Xk^{o,aux}}
\newcommand{\Xoauxh}{\Xkh^{o,aux}}
\newcommand{\Xko}[2][]{\Xk[#2]^{o#1}}
\newcommand{\Xkoh}[2][]{\Xkh[#2]^{o#1}}
\newcommand{\Xkotime}[3][]{\Xko[#1]{#2}\timenota{#3}}
\newcommand{\Xkot}[2][]{\transpose{\Xko[#1]{#2}}}
\newcommand{\Xkottime}[3][]{\Xkot[#1]{#2}\timenota{#3}}
\newcommand{\Xksh}[1]{\Xk[#1]^{\#}}
\newcommand{\Xkshtime}[2]{\Xksh{#1}\timenota{#2}}
\newcommand{\Xksht}[1]{\transpose{\Xksh{#1}}}
\newcommand{\Xkshttime}[2]{\Xksht{#1}\timenota{#2}}
\newcommand{\Lmarg}[1]{L_#1^{(marg)}}
\newcommand{\LSH}[1]{L_{#1(\#)}}
\newcommand{\muesh}[1]{\mu_{#1(\#)}}
\newcommand{\XnGen}[2]{\bm{{#1}_{b(#2)}}}
\newcommand{\Xn}[1][]{\XnGen{X}{#1}}
\newcommand{\Xntime}[2][]{\Xn[#1]\timenota{#2}}
\newcommand{\Xndst}[2]{\bm{X_{b(#1)\{#2\}}}}
\newcommand{\Xnds}[2]{\bm{X_{b(#1)[#2]}}}
\newcommand{\XndsGen}[2][\XnGen{X}{k}]{#1{}_{\bm{[#2]}}}
\newcommand{\Xndts}[2]{\bm{X_{b(#1)[\{#2\}]}}}
\newcommand{\Xnd}[2]{\bm{X_{b(#1)}\{#2\}}}
\newcommand{\Wk}[1][]{\bm{W_{#1}}}
\newcommand{\Wkh}[1][]{\hat{\bm{W}}_{\bm{#1}}}
\newcommand{\Wktime}[2][]{\Wk[#1]\timenota{#2}}
\newcommand{\Wks}[2]{\bm{W_{#1,#2}}}
\newcommand{\Wksh}[2]{\hat{\bm{W}}_{\bm{#1,#2}}}
\newcommand{\Wkstime}[3]{\Wks{#1}{#2}\timenota{#3}}
\newcommand{\Wkssh}[2]{\bm{W^\#}_{\bm{#1,#2}}}
\newcommand{\lwupgen}[4]{\{#1\}_{#2 = #3}^{#4}}
\newcommand{\lwup}[3]{\lwupgen{#1}{k}{#2}{#3}}
\newcommand{\timedur}[2]{[#1,#2]}
\newcommand{\tatb}{\timedur{\ta}{\tb}}
\newcommand{\pfull}[1]{p_{full}^{(#1)}}
\newcommand{\pfullint}[1]{p_{full}^{[#1]}}
\newcommand{\pk}[2]{p_k^{(#1,#2)}}
\newcommand{\pkm}[1]{p_k^{(#1)}}  
\newcommand{\varsetN}[3]{\{#1 \mid #1 \in \mathbb{N}, #2\leq #1\leq #3 \}}
\newcommand{\varsetnN}[2]{\{#1 \mid #1 \in \mathbb{N}, #1\neq #2 \}}
\newcommand{\unionnk}{\bigcup\limits_{\substack{i=1 \\ i\neq k}}^n}
\newcommand{\union}{\bigcup\limits_{i=1}^n}
\newcommand{\Mod}[1]{card(#1)}
\newcommand{\Modd}[1]{\left\vert{#1}\right\vert}
\DeclarePairedDelimiter\abs{\lvert}{\rvert}
\section{Introduction}\label{Introduction:section}
The state estimation of high dimensional problems is a topic of great interest in diverse research and application areas. Examples include mono and multi-agent Simultaneous Localization and Mapping problems (SLAM) \cite{SLAM1} in the robotics field; Simultaneous Structural Identification and Damage Detection of large civil structures in the structural health and monitoring (SHM) field \cite{AEKF} \cite{SystemIden} \cite{SHM}; solving Partial Differential Equations in the context of EEG estimation \cite{EEG1} \cite{EEG2} in biomedical engineering field. In all these cases, the associated estimation process needs to deal with a high dimensional state vector which implicitly requires  high dimensional multi-variate Probability Density Functions (PDF).

A traditional approach for solving such high dimensional estimation problems is the well-known Kalman Filter (KF) \cite{KF1} and its variants. KF computes the exact posterior distribution when applied to linear systems and where the sources of uncertainty behave as Gaussian white noise (GWN) \cite{KF2}. For mildly non-linear models where the estimated PDF remains symmetrical and unimodal, a variant of KF called Extended Kalman Filter (EKF) can be applied using a first order linearised model. A higher order EKF exists, such as in \cite{SOEKFtarget} and \cite{SOSEKFtarget}, but additional complexity has prohibited its widespread use \cite{PFtutorial}, particularly in high dimensional cases. The Unscented Kalman Filter (UKF), which is closely related to the second order EKF \cite{PFPositioning},  was introduced in \cite{UKF} for better approximation of the resulting PDFs, under non-linear process and observation models. The Cubature Kalman Filter (CuKF), recently introduced in \cite{arasaratnam2009cubature}, for the similar purpose of non-linear estimation, approximates the multi-variate moment integrals of Gaussian PDF by spherical radial cubature rule. The Point Mass filter (PMF) and the particle filter (PF) \cite{PF} are suited for estimating and propagating non-Gaussian distributions; however, although these  estimators are highly efficient, the required processing cost can be unaffordable, for simultaneously high-frequency high-dimensional cases.   

When an estimation problem is divided in a set of sub-problems, to reduce the processing cost and its complexity, it usually results in sacrificing optimality. That undesired effect is intensified if the operations are performed at high frequency.

The concept of ``divide and conquer'' and decentralization is the ideal way to treat such high dimensional problems. A number of such techniques are reviewed in \cite{OverviewMultiAgent}, specifically pertaining to multi-agent coordination. The attempts to decentralize and perform distributed Kalman filtering for sensor networks using consensus algorithms are given in \cite{DistributedKF1} and \cite{DistributedKF2}. The comparison of different decentralised KF schemes, including the distributed KF in \cite{DistributedKF1}, were conducted in \cite{DistributedKFHeat} for solving one dimensional linear heat conduction processes. The local estimates of the distributed KF, for time-invariant process and observation models, have been shown to converge to those of the centralised KF (full filter) with the cost of temporarily sacrificing optimality \cite{ConvergenceDistributedKF} (i.e.\  before the convergence is achieved, the estimates of the decentralized estimator differ from those of the optimal centralized one). 

Another well known approach for treating high dimensional problems is the Ensemble KF (EnKF), introduced in \cite{Ensemble1}. To avoid the computational cost associated with large number of observations in the analysis step, a conservative assumption has been made: the analysis of a certain grid point can only be impacted by the measurements which are in a neighbourhood of the said grid point \cite{Ensemble2}. This assumption is adequate for dealing with cases where the values of physical fields at distant locations are not relevantly correlated, i.e.\ the statistical dependency is localised. Although this property is common in systems modelled by PDEs, it is not a general characteristic.      
  
The GCKF was introduced in \cite{JoseGCKF} to treat certain class of problems which inherently allow the estimation process to be divided, temporarily during certain periods of time, in a subset of lower dimensional problems. In such a scenario, the compressed version of the full estimation process can be used to reduce the processing effort but still preserving optimality. The GCKF was extended from the CEKF introduced in \cite{JoseCEKF} and \cite{JoseThesis} which was intended for reducing the processing cost associated to the EKF SLAM. Additonally, the introduction of CEKF inspired the concept of virtual update  \cite{JoseCEKF}\cite{JoseThesis}, which allows for extension of CEKF to other KF variants. Without using the concept of virtual update, a recent work \cite{CUKF} has proposed a unique method for replacing the EKF core (of a CEKF), in solving mono agent SLAM problems, using the better suited UKF, for dealing with the non-linearity in the process and observation models. The GCKF also exploits the concept of stochastic cloning, orginally proposed in the CEKF in \cite{JoseCEKF} and \cite{JoseThesis}, and independently introduced for low dimensional Gaussian estimators in \cite{StochasticClone} and \cite{SC-KF}. 

This paper focuses in solving estimation processes which involve, simultaneously, high dimensionality and high frequency processing; in particularly those which are associated to systems whose dynamics are modelled through stochastic PDEs. As those systems do not satisfy the conditions required for the standard compressed estimation, approaches such as the GCKF are incapable of solving those estimation problems. The reason of this limitation is mainly due to the associated dynamic models (PDEs), which do not, at any point in space and time, decouple into a subset of lower dimensional systems.

In order to treat these estimation problems, new concepts are proposed in this work. The new approaches are the subsystem switching and new architectures for information exchange between subsystems. The joint application of these techniques allow to preserve the statistical dependency of the partitioned states, when processed in a compressed fashion.
 
Thus, similarly to those cases where compressed estimation is feasible, the proposed techniques allow treating previously intractable cases. This is achieved by preserving the statistical dependency between all the estimates, which allows achieving a comparable performance to an optimal Gaussian estimator, but at a lower processing cost.
 
The extended concepts are later applied specifically to simple SPDEs; i.e. a 1D parabolic (heat equation) and a hyperbolic (advection equation) PDEs. The results are then discussed analysing the computational benefits of the approach with respect to the full KF, in estimating a high-frequency/high-dimensional estimation process. Although the results, shown in this paper, are exclusively from heat equations; additional cases are included in the supplementary document available at \cite{WebPage}.

The paper is organised as follows. The problem statement is formulated in section~\ref{ProblemStatement:section} where the limitations of the standard GCKF and key proposals of this paper are highlighted.  The concept about the GCKF is briefly reviewed in section~\ref{GCKFreview:section}. Different information exchange architectures are proposed in section~\ref{Information:section}, to compensate the forced/artificial division of subsystems. The results are shown and discussed in section~\ref{Results:section}, where the concepts developed in section~\ref{Information:section} are applied for solving simple 1D SPDEs. Additional results including application of the proposed technique to non-linear problems, by replacing the KF core with the better suited UKF core, is also shown in section~\ref{Results:section}. A number of derivations and relevant equations with explanations are included in the appendix, for aiding and providing context to the main content.

\section{Problem Statement}\label{ProblemStatement:section}
The GCKF is limited to certain class of problems which inherently allow the estimation process to be temporarily divided, at certain intervals of time, in a subset of lower dimensional problems. In order to appreciate the concepts developed in this paper, we examine the applicability of GCKF in a low dimensional problem with increasing structural complexity.
\subsection{Decoupled Non-Overlapping Subsystem}\label{ssect:PS:DNOS}    
In this class of problems, during a certain period of time, an estimation process can be decoupled into a subset of lower dimensional subsystems. The process and observation models of the decoupled subsystems and the organisation of states can be given by equations \ref{eq:Review:estimation} and \ref{eq:Review:full_state} respectively. The division of subsystems and the organisation of states occur in a way that the estimation process has been temporarily decoupled. The individual subsystems are not allowed, during that period of time, to share common states; hence this structure (i.e.\ partition of the states) is referred to as ‘non-overlapping’. The structure and content of the subsystems may change over time, usually at low frequency, depending on the nature of the system. This process is referred to as `natural switching' (or simply `switching'); its illustration is shown in figure~\ref{img:PS:Switching}. This class of problems can be solved optimally by the GCKF described in section~\ref{GCKFreview:section}. An example of this class of estimation problem, that can be decoupled into three non-overlapping subsystems during the interval $\tatb$, is given in the equation in figure~\ref{img:PS:Switching}. It can be seen that, after a switching event at time $\tb$, during $[\tb,t_c]$, the structure is different, but still non-overlapping.
\begin{figure}[!htbp]
\centering
\begin{subfigure}[!h]{\linewidth}
\begin{equation}
 \bm{X} = \transpose{[\Xkt{1},\Xkt{2},\Xkt{3}]} = \transpose{[\transpose{\Xk[1]^*},\transpose{\Xk[2]^*}]}
 \nonumber	
\end{equation}
\end{subfigure}
\renewcommand{\tabcolsep}{0pt}
\begin{flushleft}
\begin{tabular}[!h]{@{} m{.52\linewidth}:m{.4\linewidth}@{}}
\centering 
\begin{subfigure}[h]{\linewidth}
 \centering  
 \begin{equation}
 \begin{aligned}
  \dot{\Xk[1]}\timenota{t} &= f_1(\Xktime[1]{t},t, \xi_1(t))\\
  \dot{\Xk[2]}\timenota{t} &= f_2(\Xktime[2]{t},t, \xi_2(t))\\
  \dot{\Xk[3]}\timenota{t} &= f_3(\Xktime[3]{t},t, \xi_3(t))\\
 \Xk[i]\cap & \Xk[j] = \emptyset; \forall\{ i,j\} :i\neq j \\ \forall t &\in [\ta,\tb]
 \end{aligned}
 \nonumber 
 \end{equation}
\end{subfigure}
&
\begin{subfigure}[h]{\linewidth}
 \centering  
 \begin{equation}
 \begin{aligned}
  \dot{\Xk[1]^*}\timenota{t} &= f_1^*(\Xk[1]^*\timenota{t},t, \xi_1^*(t))\\
  \dot{\Xk[2]^*}\timenota{t} &= f_2^*(\Xk[2]^*\timenota{t},t, \xi_2^*(t))\\
  \Xk[1]^* &\cap \Xk[2]^* = \emptyset \\ 
  \forall t &\in [\tb,t_c]
 \end{aligned}
 \label{eq:PS:DNOS}
 \end{equation}
\end{subfigure}
\end{tabular}
\end{flushleft}
\begin{flushleft}
\begin{tikzpicture}  
  \coordinate (A1) at ($(0,0)$) {};
  \draw ($(A1)+(0,5pt)$) -- ($(A1)-(0,5pt)$);
  \node at ($(A1)+(0,3ex)$) {$\ta$};  
  \coordinate (A2) at ($(0,0)+(0.5\linewidth,0)$) {};
  \draw ($(A2)+(0,5pt)$) -- ($(A2)-(0,5pt)$);
  \node at ($(A2)+(0,3ex)$) {$\tb$};
  \node at ($(A2)-(0,3ex)$) {($\tb$: Switching Time)};
  \coordinate (A3) at ($(0,0)+(.9\linewidth,0)$) {};
  \draw ($(A3)+(0,5pt)$) -- ($(A3)-(0,5pt)$);
  \node at ($(A3)+(0,3ex)$) {$t_c$};
  \node at ($(A3)+(.07\linewidth,0)$) {$t$};
  \draw[ultra thick, arrows=stealth-stealth] ($(A1)+(3ex,3ex)$) -- ($(A2)+(-3ex,3ex)$);
  \draw[ultra thick, arrows=stealth-stealth] ($(A2)+(3ex,3ex)$) -- ($(A3)+(-3ex,3ex)$);
  \draw[ultra thick,arrows=->] (A1) -- (A3) -- ($(A3)+(.05\linewidth,0)$);
\end{tikzpicture}
\end{flushleft}
\caption{A switching event naturally occurring at time $\tb$.}
\label{img:PS:Switching}
\end{figure}
 
\subsection{Decoupled Overlapping Subsystem}\label{ssect:PS:DOS}
A more complex structure may involve cases where individual subsystems are allowed to share some subset of states. An example can be given by modifying  equation~\ref{eq:PS:DNOS} to equation~\ref{eq:PS:DOS}. For the estimation processes in equation~\ref{eq:PS:DOS} to be decoupled, the structure of subsystems will involve sharing the subset of states $\Xk[2]$, i.e.\ the new partition will be $[\Xk[1], \Xk[2]]$ and $[\Xk[2], \Xk[3]]$. The GCKF needs to be reformulated for optimally treating this class of estimation problem. Alternatively, a decoupled overlapping subsystem can always be posed as a non-overlapping coupled subsystem.
\begin{equation}
\begin{aligned}
 \dot{\Xk[1]}\timenota{t} &= f_1(\Xktime[1]{t},\Xktime[2]{t},t, \xi_1(t))\\
 \dot{\Xk[2]}\timenota{t} &= f_2(\Xktime[2]{t},t, \xi_2(t))\\
 \dot{\Xk[3]}\timenota{t} &= f_3(\Xktime[3]{t},\Xktime[2]{t},t, \xi_3(t))
\end{aligned}
 \label{eq:PS:DOS}
\end{equation}
\subsection{Sparsely Coupled Subsystem}\label{ssect:PS:CS}
In this class of problems, an estimation process does not decouple at any point in space and time; however, it exhibits a sparse and localized structure \cite{khan2008distributing}. Examples include physical systems that result from spatio-temporal discretisation of random fields and physical phenomena characterized by a PDE \cite{khan2008distributing}. In order to exploit the computational advantages of the GCKF, the system must be temporarily partitioned into a set of lower dimensional subsystems. Since the system can not be decoupled, the division of subsystems and the `Switching' events do not longer occur naturally. The composition of the subsystems and their associated estimation processes would artificially change over time. A simplified example of this class of system is  given in equation~\ref{eq:PS:CS}.  
\begin{equation}
\begin{aligned}
\dot{\Xk[1]}\timenota{t} &= f_1(\Xktime[1]{t},\Xktime[2]{t},t,\xi_1\timenota{t}) \\ 
\dot{\Xk[2]}\timenota{t} &= f_2(\Xktime[1]{t},\Xktime[2]{t},\Xktime[3]{t},t,\xi_2\timenota{t}) \\ 
\dot{\Xk[3]}\timenota{t} &= f_3(\Xktime[2]{t},\Xktime[3]{t},\Xktime[4]{t},t,\xi_3\timenota{t}) \\ 
\dot{\Xk[4]}\timenota{t} &= f_4(\Xktime[3]{t},\Xktime[4]{t},t,\xi_4\timenota{t})
\end{aligned}
 \label{eq:PS:CS}
\end{equation}

Due to the lack of a natural partition, the individual estimation processes (or subprocesses) associated to the four subsystems in equation~\ref{eq:PS:CS} are no longer independent and do permanently require certain estimates from other subsystems. In order to partition an estimation process whose process model presents characteristics such as in equation~\ref{eq:PS:CS}, a number of techniques are applied. Some of those are for implementing the information exchange among estimation subprocesses, in an accurate and consistent way. Section \ref{Information:section} introduces novel information exchange techniques,  with the objective of applying the information from other subsystems in a manner that exploits and captures its prevailing statistical dependency to the internal states of the subsystem. 

\begin{figure}[!htb]
 \centering
 \begin{subfigure}{.49\linewidth}
  \centering
  \includegraphics[width = \linewidth]{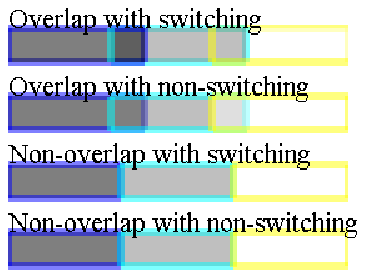}
  \caption{}
  \label{subimg:SubSysHeat21:a}
 \end{subfigure}
 \hfill
 \begin{subfigure}{.49\linewidth}
  \centering
  \includegraphics[width = \linewidth]{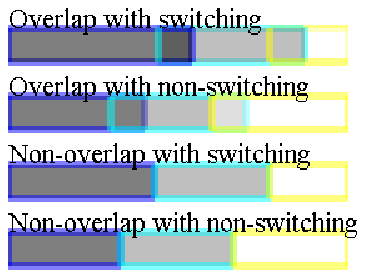}
  \caption{}
  \label{subimg:SubSysHeat21:b}
 \end{subfigure}
 \caption{Example of different subsystem archetypes (table~\ref{tab:Methods}) in a 100 dimensional system ($nos=100$) with 3 subsystems ($noss = 3$) and 10 states overlap. States of the same subsystem are shown adjacent in the full state vector, for the sake of simplicity, in this figure.}
 \label{img:SubSysHeat21}
\end{figure}

For a GCKF process, to properly approximate the operation of a full filter, two operations need to be applied: information exchange and switching. Jointly, those operations allow the GCKF process to maintain the statistical dependency among all the states' estimates in the artificially decoupled subsystems. All possible architecture for the subsystems related to switching and overlapping are shown in table~\ref{tab:Methods}. The combined system archetypes in table~\ref{tab:Methods} can be schematically visualised in figure~\ref{img:SubSysHeat21}. The subsystems that have switching capability can be seen to have different state components at different instances in time, while the subsystems without switching have a fixed structure, as shown in figure~\ref{subimg:SubSysHeat21:a} and figure~\ref{subimg:SubSysHeat21:b}. The detailed investigation of the effect of different information exchange architectures coupled with different system archetypes is given in section~\ref{Results:section}. The archetypes involving the overlapping of subsystems is outside the scope of this paper.
\begin{table}[!htb]
\caption{List of System Archetypes with their short forms} \label{tab:Methods}
\renewcommand{\arraystretch}{.5}
\begin{center}
\begin{tabular}{@{}M{.77\linewidth}M{.18\linewidth}@{}}\toprule
Combined Archetypes & Short form \\ \midrule
Non-overlapping without switching of subsystems & NONS \\ 
Non-overlapping with switching of subsystems & NOS \\ 
Overlapping without switching of subsystems & ONS \\ 
Overlapping with switching of subsystems & OS \\ \bottomrule
\end{tabular}
\end{center}
\end{table}

\section{Review of Generalized Compressed Kalman Filter (GCKF)}\label{GCKFreview:section}
Consider a case of a high dimensional estimation process which can be decoupled, during a certain period of time (for instance, during an interval $\tatb$), into a subset of lower dimensional subsystems which do evolve individually but not statistically independent, i.e.\ the systems' process and observation models are decoupled but their estimates are not independent. The estimated variable for such a process can be represented by the following state vector, which has been organised in block of states representing, in this instance, $n$ decoupled subsystems.
\begin{equation}
 \begin{matrix}
  \bm{X} = [\Xkttime{1}{t} \quad \Xkttime{2}{t} \quad \ldots \quad \Xkttime{n}{t}]^T, \\
  \Xkt{1} \in \mathbb{R}^{N_1},\ldots,\Xkt{n} \in \mathbb{R}^{N_n}
  \label{eq:Review:full_state}
 \end{matrix}
\end{equation}

Assume that the belief at time $\ta$ about the full state of the system \bm{$X$} (as given in equation~\ref{eq:Review:full_state}) at time $\ta$, $\Xktime{\ta}$, is represented by the joint PDF $\pfull{\ta}(\Xktime{\ta})$, which can be alternatively represented by $\pfull{\ta}(\lwup{\Xktime[k]{\ta}}{1}{n})$.

The full process and observation models of the decoupled subsystems, during the interval $\tatb$, are as follows:
\begin{equation}
\begin{aligned}
\left\{
\begin{array}{c}	   
 \dot{\Xk[k]}\timenota{t} = f_k(\Xktime[k]{t},u(t),t, \xi_k(t))\\ 
 \bm{Y_k}\timenota{t} = h_k(\Xktime[k]{t},t,\eta_k\timenota{t})
\end{array}
\right\}_{k=1}^n ;\forall t \in \tatb
 \end{aligned}     
 \label{eq:Review:estimation}
\end{equation}

The joint PDF of the full system at time $t$ could be represented by $\pfull{t}(\lwup{\Xktime[k]{t}}{1}{n})$. Contrary to the conventional belief, a set of lower dimensional estimation processes, individually associated to the subsystems in equation~\ref{eq:Review:estimation}, can be performed separately during the interval $\tatb$, for finally obtaining the optimal joint PDF  $\pfull{\tb}(\lwup{\Xktime[k]{\tb}}{1}{n})$, without explicitly maintaining the full PDF $\pfull{t}(\lwup{\Xktime[k]{t}}{1}{n})$ during $\tatb$. This is achieved by using the concepts of stochastic cloning and global update of the GCKF. 

The individual subsystems are augmented to represent the joint belief of the decoupled block of states along with their perfectly time invariant counter part, i.e.\ stochastic clones, such as shown for subsystem $k$ in the following,
\begin{equation}
 \Xkaugtime[k]{t} = \transpose{[\Xkttime{k}{\ta},\Xkttime{k}{t}]}
 \mathtt{\sim}\ \pk{k}{t}(\Xktime[k]{\ta}, \Xktime[k]{t})
 \label{eq:Review:stclone}
\end{equation}   

The PDF in equation~\ref{eq:Review:stclone} represents the (suboptimal) belief about the random variables $\Xktime[k]{\ta}$, $\Xktime[k]{t}$ based on the information provided by marginalizing the full PDF at time $\ta$, and the information obtained by the sequence of subsequent local predictions and observations associated   to subsystem $k$ during the period $\timedur{\ta}{t}$.
  
The individual estimation processes are modified to involve the stochastic clones of the subset of states in each subsystem; for instance the estimation of the subset of states $\Xk[k]$, associated with subsystem $k$ in equation~\ref{eq:Review:estimation}, is defined as,
\begin{equation}
\begin{array}{c}
 \varsetN{k}{1}{n} \\
 \frac{\mathrm{d}\Xkaug[k]\timenota{t}}{\mathrm{d}t} =
 \begin{cases}
  \dot{\Xk[k]}\timenota{\ta} = 0 \\
  \dot{\Xk[k]}\timenota{t} = f_k(\Xktime[k]{t}, u(t), t, \xi_k(t))
 \end{cases}
\end{array}
 \label{eq:Review:augmentedpm}
\end{equation}
\begin{equation}
 \bm{Y_k}(t) = h_k(\Xktime[k]{t}, t, \eta_k(t))
 \label{eq:Review:augmentedom}
\end{equation}

\sloppy
The prediction and observation models given in equations \ref{eq:Review:augmentedpm} and \ref{eq:Review:augmentedom} are applied during the interval $\tatb$ for finally producing $\pk{k}{\tb}(\Xktime[k]{\ta}, \Xktime[k]{\tb})$; i.e.\ the local belief about random variables $\Xktime[k]{\ta}$,$\Xktime[k]{\tb}$ at time $\tb$,

An individual virtual likelihood function, which is responsible for converting the augmented prior $\pk{k}{\ta}(\Xktime[k]{\ta}, \Xktime[k]{\ta})$ to its associated posterior $\pk{k}{\tb}(\Xktime[k]{\ta}, \Xktime[k]{\tb})$, can be synthesised and transferred to the overall joint belief at time $\ta$ to produce the optimal belief at time $\tb$ via the process called Global Update. The procedure has been represented recursively, applied for all the subsystems, as follows,
\begin{equation}
 \begin{matrix}
  \pfullint{j}(\lwup{\Xktime[k]{\tb}}{1}{j},\lwup{\Xktime[k]{\ta}}{j+1}{n}) \propto \\
  \int\limits_{\Omega_j} \pfullint{j-1}(\lwup{\Xktime[k]{\tb}}{1}{j-1},\lwup{\Xktime[k]{\ta}}{j}{n})\cdot\mu_j\cdot \\
  L_j(\Xktime[j]{\ta},\Xktime[j]{\tb})\cdot\mathrm{d}\Xktime[j]{\ta} \\
  \varsetN{j}{2}{n} 
 \end{matrix}
 \label{eq:Review:FullGB}
\end{equation} 
where, $\Omega_j$ represents the domain of $\Xktime[j]{t}$ and the factor $\mu_j$  represents the uninformative nature of the prior PDF. The likelihood function given by $L_j(\Xktime[j]{\ta},\Xktime[j]{\tb})$ in equation~\ref{eq:Review:FullGB} represents the local information collected by the subsystem $j$. The formal definition of the likelihood can be trivially evaluated using the local PDF at time $\bm{t_a}$ and $\bm{t_b}$ as given in equation~\ref{eq:Review:Likelihood}.    
 \begin{equation}
   L_k =  
   \begin{cases}
    \frac{\pk{k}{\tb}(\Xktime[k]{\ta}, \Xktime[k]{\tb})}{\pkm{\ta}(\Xktime[k]{\ta}).\mu_k},   	& \forall\Xktime[k]{\ta} : \pkm{\ta}(\Xktime[k]{\ta}) \neq 0 \\
	0, & \forall\Xktime[k]{\ta} : \pkm{\ta}(\Xktime[k]{\ta}) = 0     
   \end{cases}
  \label{eq:Review:Likelihood}
 \end{equation}

\fussy 
The application of all the individual global updates, as represented by equation~\ref{eq:Review:FullGB}, results in the joint belief of the full system at time $\tb$, $\pfull{\tb}(\lwup{\Xktime[k]{\tb}}{1}{n})$. This joint PDF is optimal and would be identical to the one obtained by maintaining the full PDF $\pfull{t}(\lwup{\Xktime[k]{t}}{1}{n})$ throughout the interval $\tatb$.

From the Gaussian perspective, the individual likelihood function is factored as the product of a constrained likelihood function  and a non-informative likelihood function, which are implemented through a virtual Gaussian observation and virtual Gaussian prediction, respectively (appendix~\ref{app:GVU}).

This concludes the main concepts about the GCKF, which are necessary for the discussion and adaptation in this paper.
 
\section{Information Exchange Architectures}\label{Information:section}
Because the partitioning of a full system, which does not satisfy equation~\ref{eq:Review:estimation}, is implemented by arbitrarily forcing it, the individual process models are not fully decoupled, which implies that their associated estimation processes are no longer temporarily independent, and do inherently require the estimates from the other subsystems (usually neighbouring or adjacent subsystems). The process models of the artificially decoupled subsystems during some interval $\tatb$ are shown in equation~\ref{eq:IEA:ProcessModel}.
\begin{equation}
 \left\{
  \dot{\Xk[k]}\timenota{t} = f_k(\Xktime[k]{t}, u(t), t, \xi_k(t),\Xntime[k]{t}) 
  \right\}_{k=1}^n 
 \label{eq:IEA:ProcessModel}
\end{equation}
where $\Xntime[k]{t}$ represents a set of states which are not estimated locally (in subprocess $k$) but are estimated by other subprocesses. These states are necessary for the process model in subsystem $k$ (as seen in equation~\ref{eq:IEA:ProcessModel}). These externally provided estimates are usually strongly correlated with the local estimates of $\Xktime[k]{t}$ and $\Xktime[k]{\ta}$. The objective of investigating different information exchange architectures is to apply this external information in a manner that  exploits and captures its prevailing statistical dependency to the internal states of the subsystem, because those externally provided estimates are not statistically independent of the local estimates.

The estimates of the states in $\Xntime[k]{t}$ are assumed, for the purposes of generality, to have been obtained from the subsystems belonging to the set $M$ with cardinality $\Mod{M}$; and the number of states in $\Xntime[k]{t}$ being $p_k$,
\begin{equation}
 \begin{aligned}
  M &= \{m_1, m_2, \ldots, m_{\Mod{M}}\};\ 
  \Xn[k] \in \{\bigcup\limits_{i=1}^{\Mod{M}} \Xk[m_i]\}; \\
  p_k &= \Mod{\Xn[k]}
  \label{eq:IEA:NeighbourStates}
 \end{aligned}
\end{equation}

In addition to forcefully partitioning the system into small subsystems, the process of switching is also artificial. The switching event is chosen to occur arbitrarily after some interval of time, at which point the system is re-partitioned into small subsystems (with different structure, as illustrated in figure~\ref{img:SubSysHeat21}). 
 
In the context of the PDEs, the process models are constructed based on the discretisation schemes which usually requires the estimates of the neighbouring states. In such a scenario, the set $M$ would usually simply contain the neighbouring subsystems; however, that is not a necessary condition for the proposed information exchange architectures.

There are numerous ways this information can be applied in the process model with varying degree of accuracy. Two main methods are proposed and explained in this section, and their performance comparison and discussion can be found in section~\ref{Results:section}.

Before the explanation of the different methods commences, the following notation is defined with close reference to equation~\ref{eq:IEA:NeighbourStates}, for the purpose of being used in representing different versions of sub-set of states:
\begin{enumerate}[label = {(\alph*)}]
\item $\XndsGen[{\Xn[k]}]{d}$ : Denotes the subset of states of $\Xn[k]$ belonging to the $d^{th}$ subsystem in $M$, i.e.\ subsystem $m_d$ with $1 \leq d \leq \Mod{M}$ .
\item $\Xndts{k}{d}$ : Denotes the subset of states of $\Xn[k]$ belonging to subsystem $d$, such that $d \in M$.
\end{enumerate}

There are two methods being considered. The first one is based on the usual assumption that those estimates are  statistically independent to those of the receiver subsystem's states. This concept is described in section~\ref{IndependentInput:ssection}. The other approach (section~\ref{FrozenStates:ssection}), which is novel and is introduced as part of this work, approximates the provided estimates, expressing them as a stochastic function of the local estimates, based on information provided by the external subsystems. This approach is remarkably superior to the usual approach presented in section~\ref{IndependentInput:ssection}.

\subsection{Independent Input}\label{IndependentInput:ssection}
This simplistic approach treats $\Xntime[k]{t}$ as an independent input to the subsystem. It ignores the statistical dependency between the input and the internal states of the subsystem, the estimates of the other states, and the input itself in previous times. This can be seen by slightly modifying the process model in equation~\ref{eq:IEA:ProcessModel} to equation~\ref{eq:IEA:II:PM}.
\begin{equation}
\begin{aligned}
 \dot{\Xk[k]}\timenota{t} &= f_k(\Xktime[k]{t},u(t),t,\xi_k(t), \XnGen{\hat{X}}{k}\timenota{t} + \XnGen{\zeta}{k}\timenota{t}) 
\end{aligned}
\label{eq:IEA:II:PM} 
\end{equation}
where, $\XnGen{\zeta}{k}$ represents an independent GWN associated with the independent input $\XnGen{\hat{X}}{k}$, whose covariance (of dimension $p_k \times p_k$, shown in equation~\ref{eq:IEA:DecorrII}) can be pessimistically decorrelated in $\Mod{M}$ subsystems using the formulation in appendix~\ref{app:Decorrelation}.
\begin{align}
 \XnGen{\bm{Q_\zeta}}{k} &= diag(\Mod{M}\cdot\bm{P_{b(k)[1]}}, \ldots, \Mod{M}\cdot\bm{P_{b(k)[p_k]}}) \nonumber\\ &= \bigoplus\limits_{i=1}^{\Mod{M}} (\Mod{M} \cdot \bm{P_{b(k)[i]}})
 \label{eq:IEA:DecorrII}
\end{align}
where, $\XndsGen[{\XnGen{\bm{P}}{k}}]{d}$ is the marginalised covariance of the set of states $\Xnds{k}{d}$ obtained from subsystem $m_d$. Thus, the additional information received in this subsystem includes the estimate and marginal belief of $\Xntime[k]{t}$, i.e.\ $\XnGen{\hat{X}}{k}$ and $\XnGen{P}{k}$.

\subsection{Exploiting local Statistical Dependency}\label{FrozenStates:ssection}
The assumption taken by the method in section~\ref{IndependentInput:ssection} that there is no statistical dependency during the interval $\tatb$, is incorrect and impractical in a large range of systems. The inconsistency of this assumption is usually severe in cases in which the approximation is applied at high frequency and in presence of highly correlated estimates. 
In addition, a relevant loss of correlation in the whole estimates is induced by it; consequently, the information collected by different subsystems is not exploited optimally.

A much more practical and realistic method is to exploit the local statistical dependency developed in the set of $M$ subsystems (the source of the estimates of $\Xn[k]$) and apply them to the local subsystem $k$. One such approach tries to achieve this  by keeping a frozen copy of the additional information taken prior to the forced subsystem partition, i.e.\ $\Xntime[k]{\ta}$, locally in the subsystem for the entire duration of $\tatb$. This requires a permanent increase in dimensionality of the local estimates from $2\times N_k$ to $2\times N_k + p_k$, demonstrated in equation~\ref{eq:IEA:FrozenStates} (modification of equation~\ref{eq:app:stclonenew}), with only an increase in dimensionality during the prediction step to $N_k + p_k$. This permanent expansion in states will allow the statistical dependency between the additional information, $\Xn[k]$, and the states in subsystem, $\Xk[k]$, to be locally developed and maintained via the constant update from the information exchange ($\Xntime[k]{t}$) between the subsystems. This particular variant of this information exchange architecture will be referred to, from henceforth, as \textit{frozen neighbours}.
\begin{equation}
 \Wktime[k]{t} =
 \transpose{[\Xkt{k}\timenota{\ta}, \Xkt{k}\timenota{t}, \transpose{\Xn[k]}\timenota{\ta}]}
 \label{eq:IEA:FrozenStates}   
\end{equation}

The additional information, $\Xntime[k]{t}$, from other subsystems can be expressed, locally in the individual $M$ subsystems, as a transformation of its time invariant counter part (stochastic clone), $\Xntime[k]{\ta}$, in a similar manner to virtual prediction in appendix~\ref{app:GVU}. The statistical approximation of $\Xntime[k]{t}$, as a function of $\Xntime[k]{\ta}$, as given in equation~\ref{eq:IEA:TransformationFrozen}, is provided by other subprocesses but exploited in the $k$ subprocess. A general expression for exploiting the local statistical dependency and the properties of the associated uncertainty is further discussed in appendix~\ref{app:GenFunc}.
\begin{equation}
 \begin{aligned}
  \Xntime[k]{t} &= G(\Xntime[k]{\ta}) +\XnGen{\bm{\zeta}}{k}\timenota{t} \\
  &= \XnGen{\bm{\alpha}}{k}\cdot(\Xntime[k]{\ta} - \XnGen{\hat{X}}{k}\timenota{\ta})\\ &{+}\: \XnGen{\hat{X}}{k}\timenota{t} + \XnGen{\bm{\zeta}}{k}\timenota{t}
  \label{eq:IEA:TransformationFrozen}
 \end{aligned}
\end{equation} 

The part of matrices $\XnGen{\bm{\alpha}}{k}$ and the statistical properties of the residual noise $\XnGen{\bm{\zeta}}{k}\timenota{t}$ (which is assumed to be GWN) can be obtained, based on the marginalised local joint PDF, $p(\Xnds{k}{d}\timenota{t}, \Xnds{k}{d}\timenota{\ta})$ in $d^{th}$ subsystem of set $M$, i.e.\ subsystem $m_d$. This is shown, in the Gaussian perspective (marginalised joint PDF is Gaussian), in equation~\ref{eq:IEA:Transformation}, in which $P$ represents the covariance matrix of states $\Wktime[m_d]{t}$ in subsystem $m_d$. It is important to note that, for times close to $\ta$, the estimates of $\Xntime[k]{t}$ and those of $\Xntime[k]{\ta}$, are strongly correlated. This strong correlation is indicated by the small covariance of the GWN $\XnGen{\bm{\zeta}}{k}\timenota{t}$, demonstrated in equation~\ref{eq:IEA:Transformation:ta}.
\begin{equation}
\renewcommand*{\arraystretch}{1}
\begin{matrix}
if, \ P(\Xnds{k}{d}\timenota{t}, \Xnds{k}{d}\timenota{\ta}) = 
\left[
 \begin{array}{;{2pt/2pt}c;{2pt/2pt}c;{2pt/2pt}}
 \hdashline \bm{A} & \transpose{\bm{C}} \\ 
 \hdashline \bm{C} & \bm{B} \\ \hdashline
 \end{array}
\right] \\
then,\ \XndsGen[{\XnGen{\bm{\alpha}}{k}}]{d} = \bm{C}\cdot\bm{A}^{-1}, \bm{Q_{\zeta_{b(k)[d]}}} = \bm{B} - \bm{C}\cdot\bm{A}^{-1}\cdot\transpose{\bm{C}}
\end{matrix}
\label{eq:IEA:Transformation} 
\end{equation}
\begin{equation}
\renewcommand*{\arraystretch}{1.0}
\begin{matrix}
\displaystyle{\lim_{t\to\ta}} P(\Xnds{k}{d}\timenota{t}, \Xnds{k}{d}\timenota{\ta}) = \left[
 \begin{array}{;{2pt/2pt}c;{2pt/2pt}c;{2pt/2pt}}
 \hdashline \bm{A} & \transpose{\bm{A}} \\ 
 \hdashline \bm{A} & \bm{A} \\ \hdashline
 \end{array}
\right]\\
\displaystyle{\lim_{t\to\ta}}\XndsGen[{\XnGen{\bm{\alpha}}{k}}]{d} = \bm{A}\cdot\bm{A}^{-1} = \bm{I},\\ \displaystyle{\lim_{t\to\ta}}\bm{Q_{\zeta_{b(k)[d]}}} = \bm{A} - \bm{A}\cdot\bm{A}^{-1}\cdot\bm{A} = \bm{0}
\end{matrix}
\label{eq:IEA:Transformation:ta} 
\end{equation}

\sloppy
In the case that the marginalised covariance, $P(\Xnds{k}{d}\timenota{\ta},\Xnds{k}{d}\timenota{\ta})$ or matrix $\bm{A}$ in equation~\ref{eq:IEA:Transformation}, is singular, a replacement for $\bm{A}^{-1}$ can be obtained via singular value decomposition or pseudo-inverse. 

The information, $\Xntime[k]{t}$, is then transferred to the subsystem by manipulating the process model to be a function of the internal frozen states, $\Xntime[k]{\ta}$, using the expression in equation~\ref{eq:IEA:TransformationFrozen}.
\begin{align}
\dot{\Wk[k]}\timenota{t} &= 
[\dot{\Xk[k]}\timenota{\ta}=0;\ \dot{\Xk[k]}\timenota{t};\ 
\XnGen{\dot{\Xk}}{k}\timenota{\ta} = 0] \\
\dot{\Xk[k]}\timenota{t} &= f_k(\Xktime[k]{t}, u(t), t, \xi_k(t), \Xntime[k]{t}) \\ &= f_k(\Xktime[k]{t}, u(t), t, \xi_k(t),
G(\Xntime[k]{\ta}) + \XnGen{\bm{\zeta}}{k}\timenota{t}) \nonumber
\end{align}

It must be noted that the Independent Input approach (in section~\ref{IndependentInput:ssection}) is a particular case of this approach, i.e.\ if it is estimated (or assumed) that no statistical dependency does exist between $\Xntime[k]{\ta}$ and $\Xntime[k]{t}$, then $\XnGen{\bm{\alpha}}{k}$ tends to be $0$ and the noise consequently increased, as in the Independent Input case. It is also worth noting that the random variable $\Xntime[k]{t}$ can also be stochastically expressed as a function of more states, if those states simultaneously belong to the local system and to other external ones; this will be subsequently discussed in this section. 

Thus, the additional information (received in this system from subsystem $m_d$) includes the transformation matrix $\bm{\alpha_{b(k)[\{m_d\}]}}$, the expected value $\bm{\hat{X}_{b(k)[\{m_d\}]}}\timenota{t}$, the expected value $\bm{\hat{X}_{b(k)[\{m_d\}]}}\timenota{\ta}$ and the covariance matrix ($\bm{Q_{\zeta_{b(k)[\{m_d\}]}}}$) of the uncertainty associated with the transformation. The construction of the matrices, $\XnGen{\bm{\alpha}}{k}$ and $\XnGen{\bm{Q_\zeta}}{k}$, using information from contributing subsystems in set $M$, requires an intuitive assumption on the arrangement of states in the frozen section of subsystem.
\begin{equation}
 \Xntime[k]{\ta} =
 \left[
 \begin{matrix}
  \Xnds{k}{1}\timenota{\ta} \\ 
  \Xnds{k}{2}\timenota{\ta} \\
  \vdots \\
  \Xnds{k}{\Mod{M}}\timenota{\ta}
 \end{matrix}
 \right]
\end{equation}

The transformation matrix, $\XnGen{\bm{\alpha}}{k}$, is a block diagonal matrix, constructed using the obtained matrices $\XndsGen[\XnGen{\bm{\alpha}}{k}]{d}$ from each subsystem in set $M$,
\begin{align}
  \XnGen{\bm{\alpha}}{k} &= diag(\bm{\alpha_{b(k)[1]}}, \bm{\alpha_{b(k)[2]}}, \ldots, \bm{\alpha_{b(k)[\Mod{M}]}})\nonumber\\ &= \bigoplus\limits_{i=1}^{\Mod{M}} \bm{\alpha_{b(k)[i]}}
\end{align}

Since the statistical dependency between the residual noise of the transformation in the different subsystems in set $M$, say, $\bm{Q_{\zeta_{b(k)[d]}}}$ and $\bm{Q_{\zeta_{b(k)[i]}}}$, is unknown, the resultant matrix, $\XnGen{\bm{Q_\zeta}}{k}$, can be pessimistically assumed to be block diagonal using the decorrelation method in appendix~\ref{app:Decorrelation}.
\begin{equation}
\bm{Q_{\XnGen{\zeta}{k}}} = \bigoplus\limits_{i=1}^{\Mod{M}} (\Mod{M}\cdot \bm{Q_{\zeta_{b(k)[i]}}})
\end{equation} 

Similarly, this concept can also be applied by keeping a frozen copy of some states (preferably the ones that would have strong correlation to $\Xntime[k]{t}$) from each individual $M$ subsystems (denoted by $\Xkotime{k}{\ta}$). The additional information, $\Xntime[k]{t}$, can be expressed, in the individual $M$ subsystems, as a function of these chosen states.
\begin{equation}
 \Xntime[k]{t} = \XnGen{\bm{\alpha}}{k}\cdot(\Xkotime{k}{\ta} - \hat{\bm{X}}_{\bm{k}}^o \timenota{\ta}) + \XnGen{\hat{X}}{k}\timenota{t} + \XnGen{\bm{\zeta}}{k}\timenota{t}
\end{equation}
where all the involved parameters ($\XnGen{\bm{\alpha}}{k}$, $\XnGen{\bm{Q_\zeta}}{k}$,  $\XnGen{\hat{X}}{k}\timenota{t}$ and $\hat{\bm{X}}_{\bm{k}}^o \timenota{\ta}$) can be synthesized in the same manner as previously discussed with the exception of the joint PDF being $p(\Xnds{k}{d}\timenota{t}, \{\Xk[m_d] \cap \Xk[k] \}\timenota{\ta})$ for the $d^{th}$ subsystem. This extension requires a change in dimensionality of the prediction step from $N_k + p_k$ to $N_k + \Mod{\Xko{k}}$. 

Any information for estimating the statistical dependency between estimates of $\Xntime[k]{t}$ and $\Xktime[k]{t}$ is useful for the individual estimation process in susbystem $k$. Therefore, the combination of the previously discussed methods, i.e.\ using the chosen frozen states in combination with the frozen additional information, may be useful. 
 
\section{Experiment and Results}\label{Results:section}
One of the main techniques, proposed for maintaining the statistical dependency in the artificially decoupled subsystem, is the Switching of Subsystems. The performance of this method, under different information exchange architectures, is shown in this section; specifically in solving the one dimensional heat PDE. The list and short forms for the information exchange architectures are given in table~\ref{tab:IEA}.  

\begin{table}[!htb]
\caption{List of Information Exchange Architectures with their short forms} 
\label{tab:IEA}
\renewcommand{\arraystretch}{.5}
\begin{tabular}{@{}M{.7\linewidth}M{.25\linewidth}@{}}\toprule
Information Exchange Architecture & Short form \\ \midrule
Independent Input & II \\ 
Exploiting local Statistical Dependency using frozen neighbours & ELSD-FN \\ 
Exploiting local Statistical Dependency using frozen chosen states & ELSD-FC \\ \bottomrule
\end{tabular}
\end{table}

The experiment has been conducted on the one dimensional heat PDE shown in  equation~\ref{eq:Results:PDEs:heat}. The simulated output used in the experiment for heat equation was obtained by adding white Gaussian noise to the numerical solution of equation~\ref{eq:Results:PDEs:heat}. The other details about the PDE, such as the chosen discretisation schemes, the initial condition and the associated parameters (used in equation \ref{eq:Results:PDEs:heat}) are given in the appendix~\ref{app:PDEs}. 
\begin{equation}
   \frac{\partial U(x,t)}{\partial t} = \beta \frac{\partial^2 U(x,t)}{\partial x^2} 
\label{eq:Results:PDEs:heat} 
\end{equation}

The results, which are obtained from the different combinations, are compared against each other in their ability to mimic  and reproduce those of the full filter. A method is considered acceptable only if its estimates (expected value and covariance matrix) are on par to those of the full filter. The ability of the compressed filter to capture the prevalent statistical dependency is verified by visually comparing its normalised covariance matrix against the full filter. The normalised covariance matrix is simply a cross correlation matrix containing Pearson's correlation coefficients, given by $\abs{cov(x,y)}/(\sigma_x \sigma_y)$ for each possible pair of scalar random variables $x$ and $y$. The gray levels in these plots signify the degree of correlation. A comparison of the standard deviations will also be conducted, where the percentage difference of various methods against the full filter will be graphically displayed. The consistency of the expected value is verified through a metric called `average discrepancy', given by the average of $\abs{\hat{\bm{X}}_{full}\timenota{t} - GT\timenota{t}} - \abs{\hat{\bm{X}}_{GCKF}\timenota{t} - GT\timenota{t}}$ in $250$ independent Monte Carlo runs (initialised with the same condition), for all times $t$ sampled at global update frequency. The details about the method used for generating the ground truth (GT) data (i.e.\ the solution of the PDE) can be found in appendix~\ref{app:PDEs}.

Additionally, the method must also be capable of achieving the same performance when dealing with PDEs with fast and slow dynamics. This can be tested by varying the parameter $s$ (i.e.\ the parameters that are responsible for $s$) in the interested PDEs given in appendix~\ref{app:PDEs}. In such a scenario, the compressed version of the full filter would be largely preferred due to its relevant saving in processing cost, and low departure from optimality.  A summary table for the parameters used in different experiments (specified under different experiment numbers) in this section is provided in table~\ref{tab:SumParameters}. The parameters that are not mentioned in the tables are constant in the experiments ($pf = 100$, $uf = 100$, $guf = 1$, $nos = 500$, $nof = 4$, $noss = 10$, $\sigma^2_o = 10$, $\sigma^2_s = 10^{-9}$, $k = 400$, $\rho = 8700$, $C_p = 385$). The notations are in accordance to table~\ref{tab:Parameters}.
\begin{table}[!hbt]
\caption{Summary list of parameters in the conducted experiments for solving one dimensional heat SPDE}
\label{tab:SumParameters}
\renewcommand{\arraystretch}{.5}
\begin{tabular}{@{}M{.08\linewidth}M{.18\linewidth}M{0.08\linewidth}M{0.18\linewidth}M{0.35\linewidth}@{}} \toprule
Exp. No. & Figures   & $l$  & System Dynamic & Output Locations \\ \midrule
1 & \ref{img:IndependentInputHeatSPV25} & 15 & $0.0013$ & [1,250,500]   \\ 
2 & \ref{img:IndependentInputHeatSPV41:NONS},\ref{img:IndependentInputHeatSPV41:NOS} & 15 & $0.0013$ & pure predictions \\ 
3 & \ref{img:IndependentInputHeatSPV33:NONS}, \ref{img:ExploitingLocalHeatSTDP33},\ref{img:ExploitingLocalHeatSPV33},\ref{img:ExploitingLocalHeatSTDP56} & 1 & $0.2997$ & pure predictions\\ 
4 & \ref{img:ExploitingLocalHeatDB44} & 1 & $0.2997$ & [1,100,200,300,400,500]\\ \bottomrule
\end{tabular} 
\end{table}
\begin{table}[!htpb]
\renewcommand{\arraystretch}{.5}
\caption{List of Parameters with their short forms}\label{tab:Parameters}
\begin{tabular}{@{}ccc@{}}\toprule
{Parameters} & {Units} & {Short form} \\ \midrule
Prediction Frequency & \SI{}{\hertz} & pf \\ 
Update Frequency & \SI{}{\hertz} & uf \\ 
Global Update Frequency  & \SI{}{\hertz} & guf \\ 
Switching Frequency & \SI{}{\hertz} & sf \\ 
Number of States & - & nos \\ 
Number of Subsystems & - & noss \\ 
Number of chosen frozen States & - & noc \\ 
Output Locations & - & ol \\ 
Variance of output & \SI{}{\celsius^2} & $\sigma^2_o$  \\ 
Variance of system & \SI{}{\celsius^2} & $\sigma^2_s$ \\ 
Thermal Conductivity & \SI{}{\watt\per\meter\per\kelvin} & k\\ 
Density & \SI{}{\kilogram\per\metre^3} & $\rho$ \\ 
Specific Heat & \SI{}{\watt\second\per{\kilogram\celsius}}  & $C_p$ \\ 
Length & \SI{}{\metre} & l \\ \bottomrule
\end{tabular}
\end{table}

In order to study and verify the performance of the GCKF in maintaining the statistical dependency of the states' estimates among different subsystems, the initial covariance is assumed to follow an exponential pattern with the formula specified in equation~\ref{eq:Results:InitiCov}.
\begin{equation}
 \begin{matrix}  
  P(i,j) = scale\times e^{-\Modd{i-j}/\phi}, \forall i \neq j \\
  P(i,i) =  scale + \psi
 \end{matrix}
 \label{eq:Results:InitiCov}
\end{equation}

This section is divided in a similar manner as section~\ref{Information:section} such that the result in each section corresponds to a particular version of GCKF (i.e.\ the GCKF under the combination of one of the system archetypes outlined in table~\ref{tab:Methods} and one of the proposed information exchange architectures). The results from the heat PDE is shown in subsections \ref{ssect:Results:II} - \ref{ssect:Results:ELSD}. Finally, the comparisons of processing cost of the full filter and the different versions of GCKF are shown and discussed in section~\ref{ssect:Results:ProcessingCost}. Briefly, some results from replacing the KF core with UKF core is shown for solving non-linear inviscid Burgers equation in section~\ref{ssect:Results:UKF}. 

\subsection{Independent Input (II)}\label{ssect:Results:II}
\begin{figure}[!htb]
 \centering
 \begin{subfigure}[t]{.25\linewidth}
  \centering
  \includegraphics[width = \linewidth, keepaspectratio]{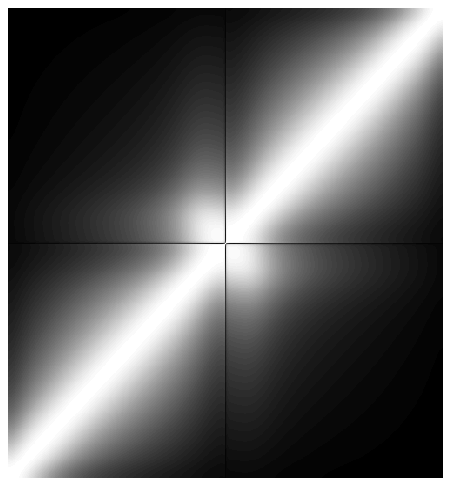}
  \caption{Full Filter} 
  \label{subimg:IndependentInputHeatSPV25:full}
 \end{subfigure}
 \begin{subfigure}[t]{.25\linewidth}
  \centering
  \includegraphics[width = \linewidth, keepaspectratio]{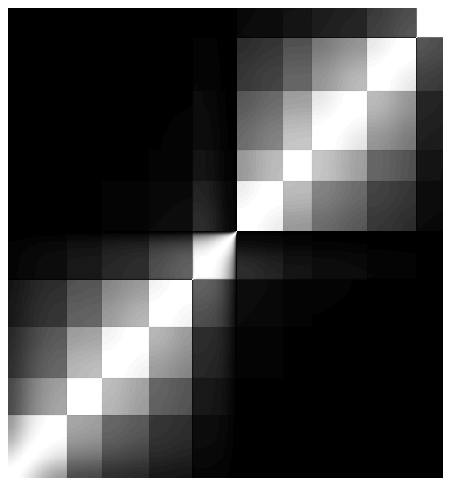}
  \caption{NONS}
  \label{subimg:IndependentInputHeatSPV25:nons}
 \end{subfigure}
 \begin{subfigure}[t]{.25\linewidth}
  \centering
  \includegraphics[width = \linewidth, keepaspectratio]{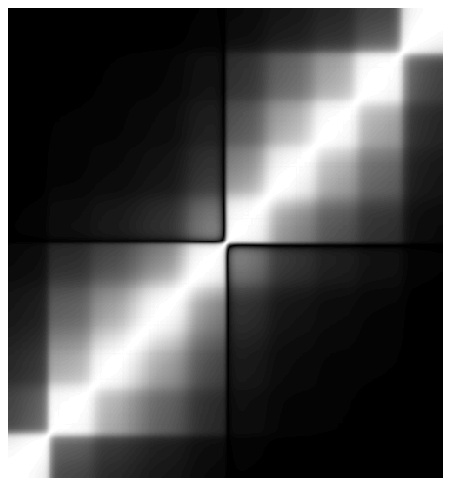}
  \caption{NOS}
  \label{subimg:IndependentInputHeatSPV25:nos}
 \end{subfigure}
 \caption{Comparison of individually normalised covariance matrices at the end of the simulation between different variants of GCKF (using II information exchange architecture) against the full filter. In the images, the regions with lighter shades of gray indicate strong correlation (or anti-correlation) while the regions with darker shades of gray indicate weak correlation between the random variables.}
 \label{img:IndependentInputHeatSPV25}
\end{figure}

The performance of the GCKF, during all the estimation process, in capturing the statistical dependency, using the archetypes outlined in table~\ref{tab:Methods}, coupled with Independent Input (II) information exchange architecture, for the one dimensional heat equation with certain parameters (specified under experiment 1), representing a slow dynamic system, is shown in figure~\ref{img:IndependentInputHeatSPV25}. 

It can be seen that none of the system archetypes (NONS and NOS) using this information exchange were able to maintain the statistical dependency between the states' estimates (in the same manner as the full filter). This can be clearly seen by comparing the normalised covariance matrix of the GCKF using NONS and NOS in figure~\ref{subimg:IndependentInputHeatSPV25:nons} and  figure~\ref{subimg:IndependentInputHeatSPV25:nos} respectively to that of the full filter in figure~\ref{subimg:IndependentInputHeatSPV25:full}. This means that information provided by subsequent observations, would not be properly exploited by the whole process.
 
\begin{figure}[!htbp]
 \centering
 \setlength{\tabcolsep}{0pt}
 \begin{tabular}{|M{.25\linewidth}|M{.25\linewidth}|M{.25\linewidth}|}
 \hline
 NoGU = 10& NoGU = 50& NoGU = 200\\
 \end{tabular}
 \begin{subfigure}{\linewidth}
  \centering
  \includegraphics[width = .25\linewidth]{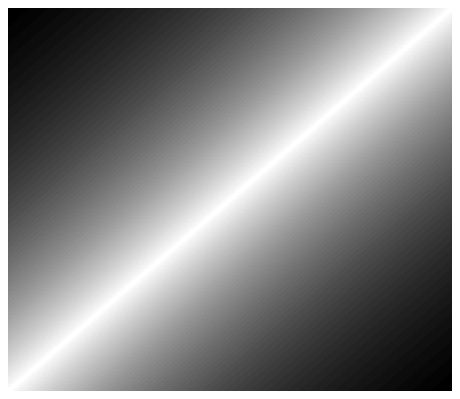}
  \includegraphics[width = .25\linewidth]{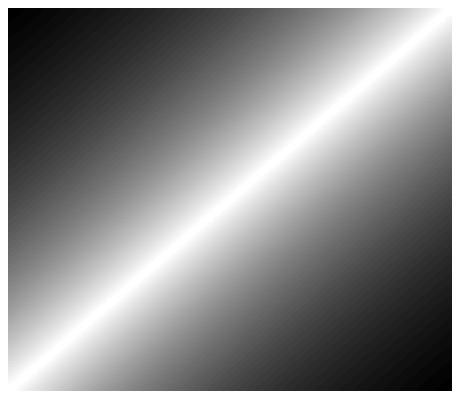}
  \includegraphics[width = .25\linewidth]{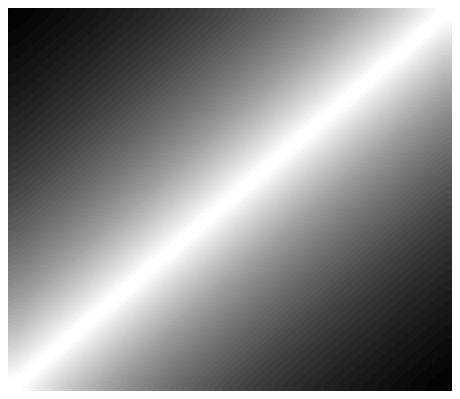}
  \caption{Full Filter}
  \label{subimg:IndependentInputHeatSPV41:NONS:full}
 \end{subfigure}
 \par
 \begin{subfigure}{\linewidth}
  \centering
  \includegraphics[width = .25\linewidth]{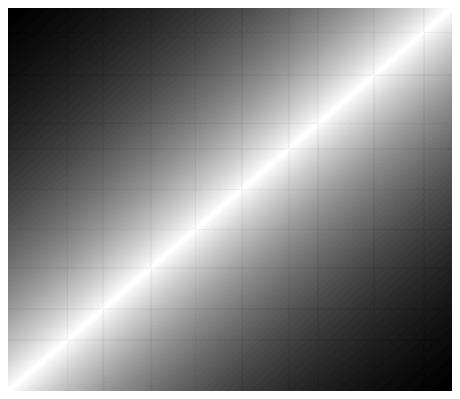}  
  \includegraphics[width = .25\linewidth]{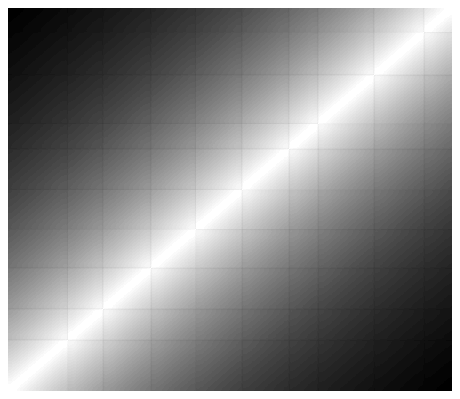}  
  \includegraphics[width = .25\linewidth]{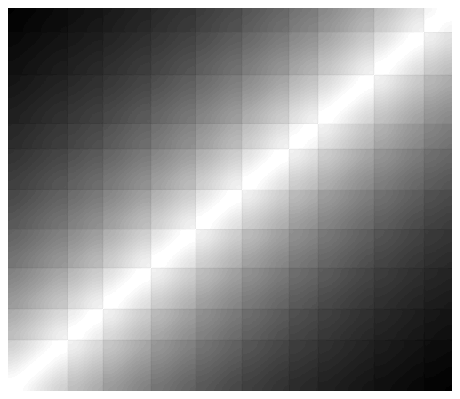}
  \caption{GCKF with NONS/II}
  \label{subimg:IndependentInputHeatSPV41:NONS:compressed}
 \end{subfigure}
 \caption{Comparison of individually normalised covariance matrices in experiment number 2 between \subref{subimg:IndependentInputHeatSPV41:NONS:full} full filter and the \subref{subimg:IndependentInputHeatSPV41:NONS:compressed} GCKF using II information exchange architecture with NONS archetype (NONS/II) at the $10^{th}$, $50^{th}$ and $200^{th}$ global update (left to right). `NoGU' abbreviates the number of global updates.} 
 \label{img:IndependentInputHeatSPV41:NONS}
\end{figure}

\begin{figure}[!htbp]
 \centering
 \setlength{\tabcolsep}{0pt}
 \begin{tabular}{|M{.45\linewidth}|M{.45\linewidth}|}
 \hline
 Full Filter & GCKF with NONS/II \\
 \end{tabular}
 \begin{subfigure}{\linewidth}
  \centering
  \begin{minipage}{.45\linewidth}
   \centering
  \includegraphics[width = .56\linewidth, keepaspectratio]{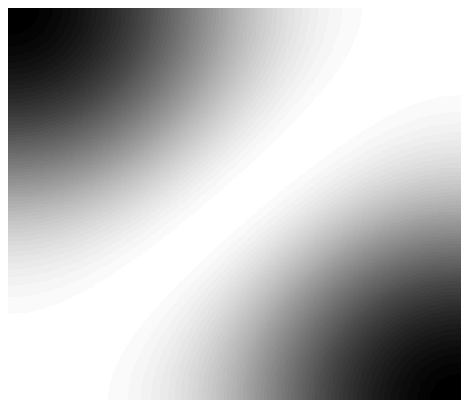}
  \end{minipage}
  \begin{minipage}{.45\linewidth}
   \centering
  \includegraphics[width = .56\linewidth, keepaspectratio]{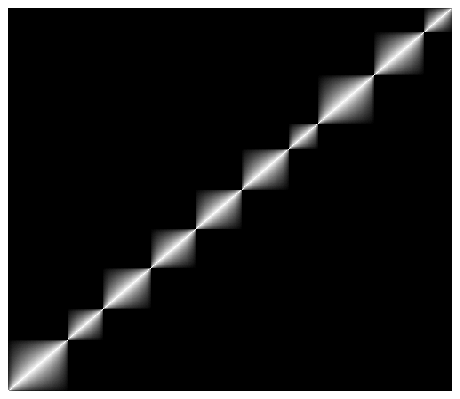}
  \end{minipage}
 \end{subfigure}
 \caption{Comparison of individually normalised covariance matrices in experiment number 3 between the full filter (left column) and the GCKF with NONS/II (right column) at the  $200^{th}$ global update.}
 \label{img:IndependentInputHeatSPV33:NONS}  
\end{figure}

The performances of the archetype that does not involve switching of subsystems, i.e.\ NONS, is expected to deteriorate over time due to its inherent inability in maintaining the statistical dependencies between the internal and external states of the subsystem. This is clearly demonstrated, as an example for the NONS archetype, by the (undesirable) presence of low cross-correlation lines between the subsystems in the normalised covariance matrices resulting in blocks of monochromatic shade in figure~\ref{subimg:IndependentInputHeatSPV41:NONS:compressed}. The statistical dependencies between the subsystems are represented by the blocks that are located off-diagonal in the normalised covariance plot (as shown in figure~\ref{subimg:IndependentInputHeatSPV41:NONS:compressed}). These blocks will slowly darken and finally turn completely black (i.e.\ zero cross-covariance), when the statistical dependencies between the subsystems' estimates are completely lost and the subsystems turn to be statistically independent. This phenomenon is not visible in the duration of the experiment conducted in figure~\ref{img:IndependentInputHeatSPV41:NONS} (experiment 2) since the simulated system had a slow dynamics, i.e.\ the rate of evolution of the statistical dependency between the states is slow. In another experiment, where the PDE parameters correspond to a system with faster dynamics, the loss of statistical dependency is apparent and occurred as early as 200 global updates, shown in  figure~\ref{img:IndependentInputHeatSPV33:NONS} (experiment 3). The gradual loss of statistical dependency is responsible for the discrepancy between the optimal (i.e\ from the full filter) marginalised standard deviations and the approximated ones (GCKF), as will be seen in  the experiment in figure~\ref{subimg:ExploitingLocalHeatSTDP33:NOS/II}, where the discrepancy almost reaches $100\%$. 

\begin{figure}[!htbp]
 \centering
 \setlength{\tabcolsep}{0pt}
 \begin{tabular}{|M{.05\linewidth}|M{.34\linewidth}|M{.34\linewidth}|}
 \cline{2-3}
 \multicolumn{1}{c|}{} & Full Filter & GCKF with NOS/II \\ \hline
 \begin{turn}{90} NoGU = 100 \end{turn} &
  \begin{subfigure}{\linewidth}
   \centering
   \includegraphics[width = 1.0\linewidth, keepaspectratio]{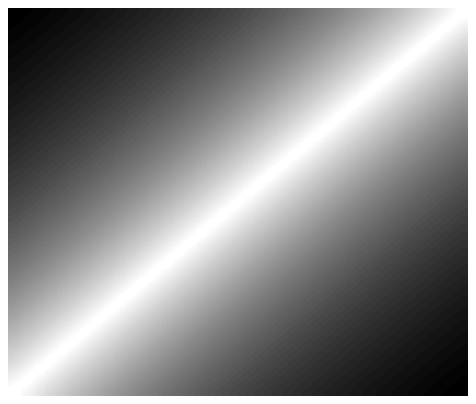}
  \end{subfigure} 
  &
  \begin{subfigure}{\linewidth}
   \centering
   \includegraphics[width = 1.0\linewidth, keepaspectratio]{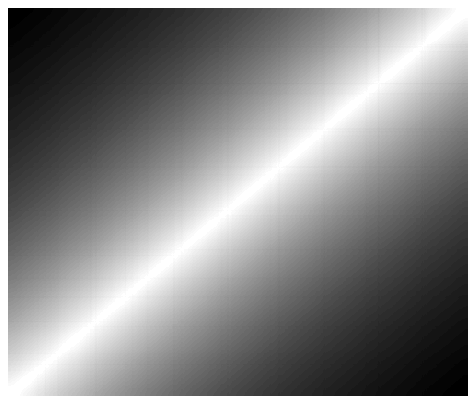}
  \end{subfigure} 
  \\ \hline
  \begin{turn}{90} NoGU = 700 \end{turn} &
  \begin{subfigure}{\linewidth}
   \centering
   \includegraphics[width = 1.0\linewidth, keepaspectratio]{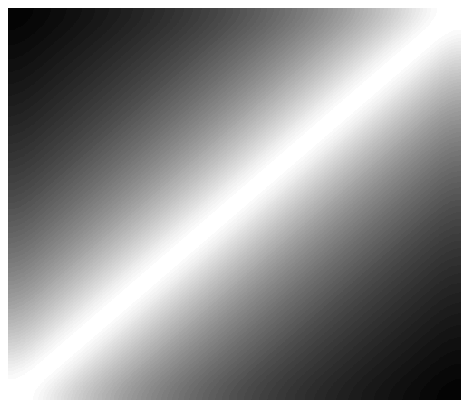} 
  \end{subfigure}
  &
  \begin{subfigure}{\linewidth}
   \centering
   \includegraphics[width = 1.0\linewidth, keepaspectratio]{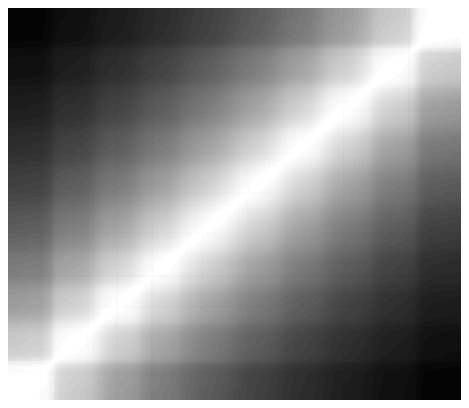}
  \end{subfigure} 
   \\ \hline
 \end{tabular}
  \caption{Comparison of individually normalised covariance matrices between the full filter (left) and the GCKF with NOS/II (right) at the $100^{th}$ and $700^{th}$ global update.}
  \label{img:IndependentInputHeatSPV41:NOS}
\end{figure}

The NOS archetype, in this instance (with II information exchange architecture, can also be referred to by NOS/II), was inadequate in retaining the statistical dependency between the subsystems for the duration of the experiment. This is demonstrated by examining the evolution of the normalised covariance matrix of the NOS archetype, as shown in figure~\ref{img:IndependentInputHeatSPV41:NOS}. The loss of statistical dependency occurs at a much slower rate as compared to the NONS archetype, where visible dark lines between the subsystems are already formed before the $200^{th}$ global update, as shown in the rightmost column of figure~\ref{subimg:IndependentInputHeatSPV41:NONS:compressed}, as opposed to the indistinct lines formed after the $700^{th}$ global update in the NOS archetype shown in the bottom right picture of figure~\ref{img:IndependentInputHeatSPV41:NOS}.
\subsection{Exploiting Local Statistical Dependency (ELSD)}\label{ssect:Results:ELSD}
\begin{figure}[!htbp]
 \centering
 \begin{subfigure}{.49\linewidth}
  \centering
  \includegraphics[width = \linewidth, keepaspectratio]{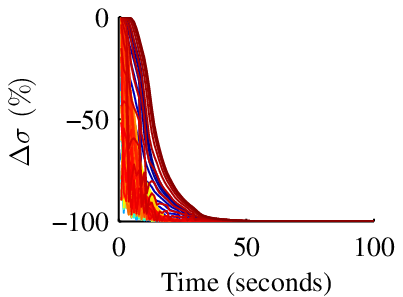}
  \caption{GCKF + NOS/II}
  \label{subimg:ExploitingLocalHeatSTDP33:NOS/II} 
 \end{subfigure}
 \begin{subfigure}{.49\linewidth}
  \centering
  \includegraphics[width = \linewidth, keepaspectratio]{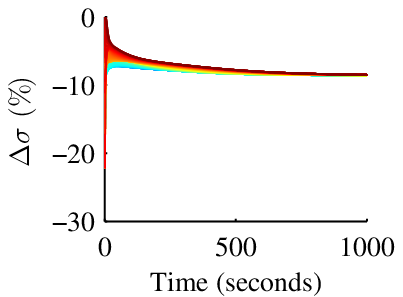}
  \caption{GCKF + NOS/ELSD-FN}
  \label{subimg:ExploitingLocalHeatSTDP33:NOS/ELSD-FN} 
 \end{subfigure}
 \caption{Comparison of percentage difference in standard deviations of the GCKF with switching archetype (NOS) and the full filter. GCKF using II information exchange architecture is shown in the left column while the GCKF using ELSD-FN information exchange architecture is shown on the right.}
 \label{img:ExploitingLocalHeatSTDP33} 
\end{figure}

The GCKF exploiting locally developed statistical dependencies (ELSD) is the only method where the estimates from the other subsystems ($\Xntime[k]{t}$) are correctly and statistically applied, and is, thus, expected to outperform the II information exchange architecture. The comparison of the performance of this method to the II architecture coupled with switching archetype (NOS) is shown in figure~\ref{img:ExploitingLocalHeatSTDP33}.
\begin{figure}[!htbp]
 \centering
 \begin{subfigure}{.49\linewidth}
  \centering 
  \includegraphics[width = \linewidth, keepaspectratio]{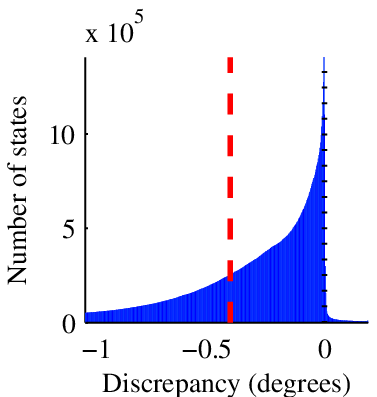}
  \caption{GCKF + NOS/II}
  \label{subimg:ExploitingLocalHeatDB44:NOS/II} 
 \end{subfigure}
 \begin{subfigure}{.49\linewidth}
  \centering 
  \includegraphics[width = \linewidth, keepaspectratio]{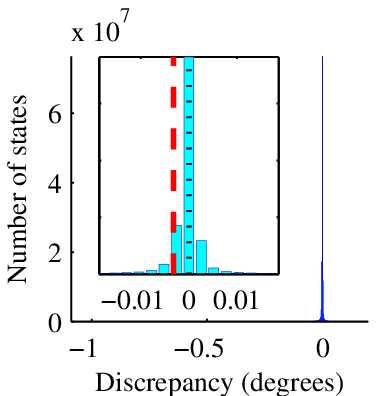}
  \caption{GCKF + NOS/ELSD-FN}
  \label{subimg:ExploitingLocalHeatDB44:NOS/ELSD-FN} 
 \end{subfigure}
 \caption{Comparison of expected temperatures (given by $\abs{\hat{\bm{X}}_{full} - GT} - \abs{\hat{\bm{X}}_{GCKF} - GT}$. The black dotted line represents no discrepancy, while the red dashed line represents the average deviation of the expected values. Positive values indicate that the discrepancy of the GCKF's expected values is lower than that of the full filter. GCKF using II information exchange architecture is shown on the left while the GCKF using ELSD-FN information exchange architecture is shown on the right.}
 \label{img:ExploitingLocalHeatDB44}
\end{figure}

\begin{figure}[!htbp]
 \centering 
 \setlength{\tabcolsep}{0pt}
 \begin{tabular}{|M{.05\linewidth}|M{.3\linewidth}|M{.3\linewidth}|M{.3\linewidth}|}
 \cline{2-4}
 \multicolumn{1}{c|}{} & Full Filter& GCKF with NOS/II& GCKF with NOS/ELSD-FN \\ \hline
 \begin{turn}{90} NoGU = 10 \end{turn} &
 \begin{subfigure}{\linewidth}
  \centering
  \includegraphics[width = \linewidth, height = .8\linewidth]{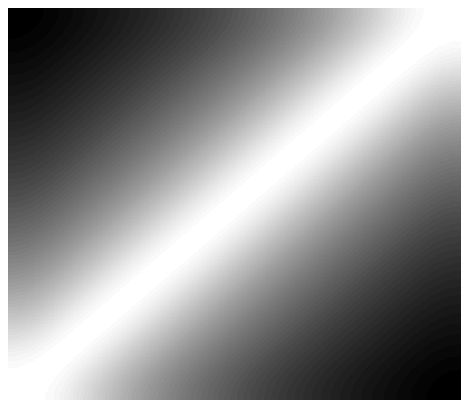}
  \end{subfigure} &
  \begin{subfigure}{\linewidth}
  \centering
  \includegraphics[width = \linewidth, height = .8\linewidth]{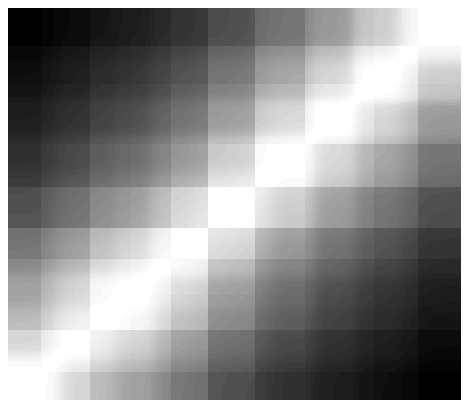}
  \end{subfigure} &
  \begin{subfigure}{\linewidth}
  \centering
  \includegraphics[width = \linewidth, height = .8\linewidth]{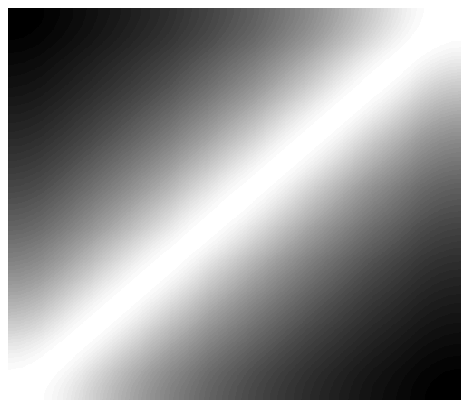}
  \end{subfigure} 
  \\ \hline
 \begin{turn}{90} NoGU = 50 \end{turn} & 
 \begin{subfigure}{\linewidth}
  \centering
  \includegraphics[width = \linewidth, height = .8\linewidth]{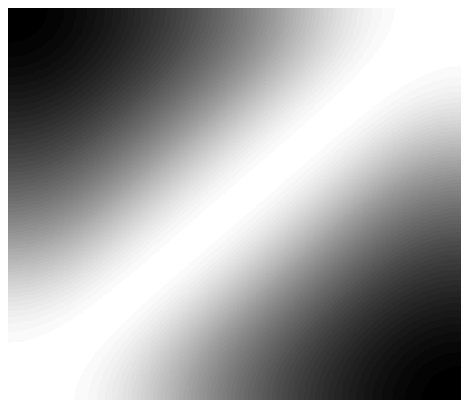}
 \end{subfigure} &
 \begin{subfigure}{\linewidth}
  \centering
  \includegraphics[width = \linewidth, height = .8\linewidth]{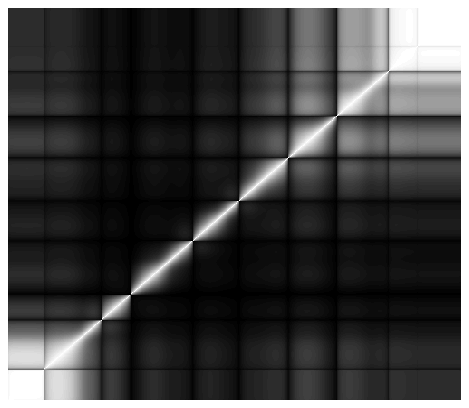}
 \end{subfigure} &
 \begin{subfigure}{\linewidth}
  \centering
  \includegraphics[width = \linewidth, height = .8\linewidth]{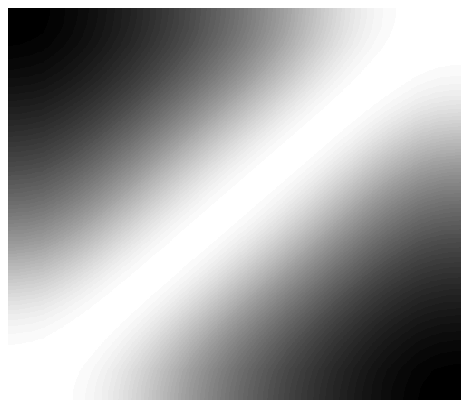} 
 \end{subfigure}
 \\ \hline
 \begin{turn}{90} NoGU = 100 \end{turn} & 
 \begin{subfigure}{\linewidth}
  \centering
  \includegraphics[width = \linewidth, height = .8\linewidth]{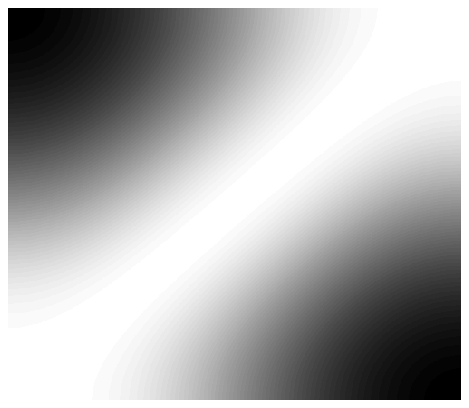}
  \end{subfigure} &
 \begin{subfigure}{\linewidth}
  \centering
  \includegraphics[width = \linewidth, height = .8\linewidth]{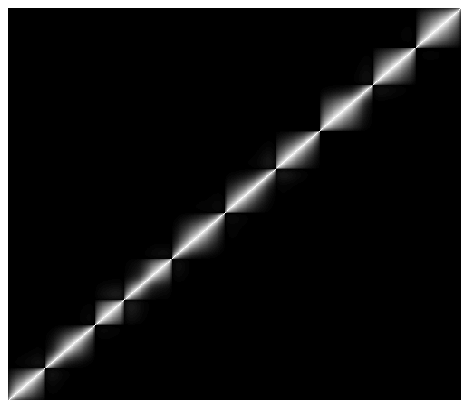}
  \end{subfigure} &
 \begin{subfigure}{\linewidth}
  \centering
  \includegraphics[width = \linewidth, height = .8\linewidth]{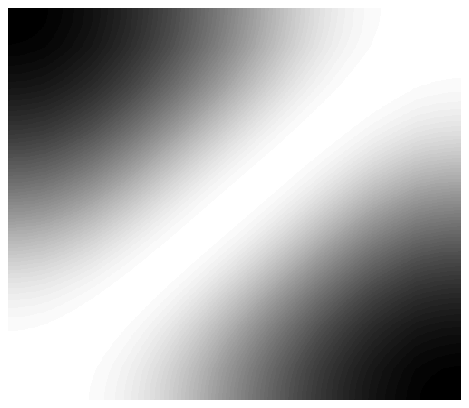}  
 \end{subfigure} \\ \hline
 \end{tabular}    
 \caption{Comparison of individually normalised covariance matrices between the full filter and variants of GCKF at $10^{th}$, $50^{th}$ and $100^{th}$ global update. The variants of filter, given from the leftmost to rightmost columns, are: full filter; GCKF with NOS/II; GCKF with NOS/ELSD-FN.}
 \label{img:ExploitingLocalHeatSPV33}
\end{figure}

The comparison conducted in figure~\ref{img:ExploitingLocalHeatSTDP33} corresponds to the case of a PDE system with extremely fast dynamics. The switching system archetype coupled with ELSD-FN information exchange architecture outperforms all the other methods by having the smallest percentage error (the only method that has less than $100\%$ error) in the marginalised standard deviations when compared against the full filter (right column of figure~\ref{img:ExploitingLocalHeatSTDP33}). Similarly, the improvement in performance can also be observed in the expected values, where the average discrepancy, indicated by the dashed red line, of NOS/ELSD-FN (figure~\ref{subimg:ExploitingLocalHeatDB44:NOS/ELSD-FN}) is lower than that of the NOS/II version of GCKF (figure~\ref{subimg:ExploitingLocalHeatDB44:NOS/II}) by order of magnitudes ($-0.4292 \SI{}{\celsius}$ against $-0.0032\SI{}{\celsius}$). The bar graph (in figure~\ref{img:ExploitingLocalHeatDB44}) shows the discrepancy of the absolute errors (between the expected temperatures and the simulated ones) for the full filter and the GCKF. The capability of this information exchange in maintaining the statistical dependency similarly to the full filter in a system with fast dynamics is shown in figure~\ref{img:ExploitingLocalHeatSPV33}, where the resemblance of the normalised covariance matrices at various global updates is observed.

The GCKF with II information exchange architecture (shown in the second column for NOS archetype, in figure~\ref{img:ExploitingLocalHeatSPV33}) lost the statistical dependencies completely as early as after the first 100 global updates; however, the GCKF using switching archetype with ELSD-FN information exchange architecture (shown in the third column), was able to keep track of the statistical dependency, similarly to the full filter (first column).

\begin{figure}[!htbp]
 \centering
 \begin{subfigure}{.49\linewidth}
  \centering
  \includegraphics[width = \linewidth, keepaspectratio]{STDCoComparePer_d-500_batch-33_variant-12}
  \caption{GCKF with ELSD-FN}
  \label{subimg:ExploitingLocalHeatSTDP56:FN}
 \end{subfigure}
 \begin{subfigure}{.49\linewidth}
  \centering
  \includegraphics[width = \linewidth, keepaspectratio]{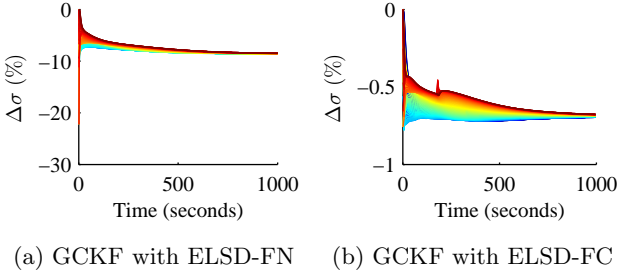}
  \caption{GCKF with ELSD-FC}
  \label{subimg:ExploitingLocalHeatSTDP56:FC}
 \end{subfigure}
 \caption{Comparison of percentage difference in standard deviations of the GCKF with NOS archetype and ELSD information exchange architecture and the full filter.}
 \label{img:ExploitingLocalHeatSTDP56}  
\end{figure}

A comparison of the performance of the GCKF with different variants of ELSD coupled with NOS system archetype is shown in figure~\ref{img:ExploitingLocalHeatSTDP56}. There is a significant performance improvement using ELSD-FC as compared to ELSD-FN. In a scenario where the system has slow dynamics, the difference in performance will be negligible and the variant can be chosen according to the associated processing cost, discussed in section~\ref{ssect:Results:ProcessingCost}.

The GCKF with NOS archetype coupled with ELSD information exchange architecture can be concluded to be the only versions that are capable of achieving good performance in slow and fast dynamic systems, and are suitable replacements for the full filter in solving this type of PDE. The results from other PDEs can be found in the supplementary document.
\subsection{Comparison of Processing Times}\label{ssect:Results:ProcessingCost}
The performance of the GCKF, with different information exchange architectures and system archetypes, in mimicking the full filter, is thoroughly investigated in this paper for evaluating the computational advantages associated with the GCKF. In order to quantitatively measure the computational cost reduction in the GCKF, when compared against the full filter, a metric called \textit{`cost ratio'} (CR) is defined in equation~\ref{eq:Results:CostRatio}. This metric represents the ratio of the cost associated with the full filter to the cost associated with the GCKF under the assumption that the prediction frequency ($pf$) is the same as the update frequency ($uf$). 
\begin{equation}
 CR = \frac{t_{ff}\cdot pf}{t_{gckf}\cdot pf + t_{gu}\cdot guf} = \frac{t_{ff}}{t_{gckf} + t_{gu}\cdot \frac{guf}{pf}}
 \label{eq:Results:CostRatio}
\end{equation}
where, the timing variable $t_{ff}$ refers to the time taken by the full filter while the variable $t_{gckf}$ refers to the time taken by the GCKF, both measured from a single iteration of prediction followed by update.

\begin{figure}[!htbp]
 \centering
 \includegraphics[width = .7\linewidth, keepaspectratio]{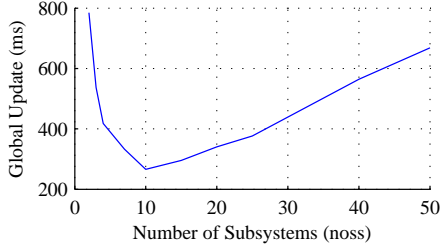}
 \caption{Comparison of the cost of global update $t_{gu}$ against the number of subsystems. The measured time is in the unit of milliseconds. The timing has been collected with the following fixed parameters: $nos = 1000$; $M = 500$; $noc = 4$.}
 \label{img:TimeComparisons:GuCost}
\end{figure}

\begin{figure}[!htbp]
 \centering
 \includegraphics[width = .9\linewidth, keepaspectratio]{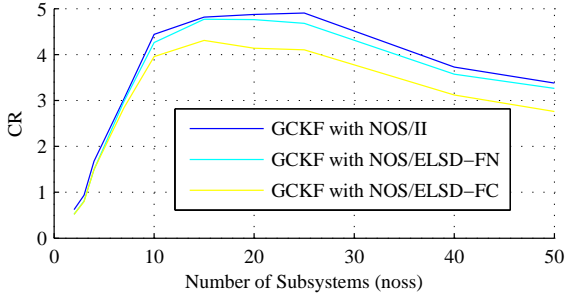}
 \caption{Comparison of CR of GCKF with different information exchange architectures and NOS archetype. The CR has been collected with same parameters as figure~\ref{img:TimeComparisons:GuCost}.}
 \label{img:TimeComparisons:CostRatio:NOSS}
\end{figure}

The dominant cost of the GCKF is the global update, $t_{gu}$ in equation~\ref{eq:Results:CostRatio}, which is approximately proportional to equation~\ref{eq:Results:CostGU}. The effect of the number of subsystems on the cost of the global update, assuming that the states and the observations are uniformly divided in each subsystem, is shown graphically in figure~\ref{img:TimeComparisons:GuCost}. The high value of $t_{gu}$, observed for systems with smaller number of subsystems ($noss < 10$), in figure~\ref{img:TimeComparisons:GuCost}, is due to the dominance of the cost associated with factor $\sum_{i=1}^n N_i^3$ in equation~\ref{eq:Results:CostGU} (such as matrix multiplication and singular value decomposition). On the other hand, the large values of $t_{gu}$ observed for systems with large number of subsystems ($noss > 10$), is due to the increase in dominance of the cost associated with factor $\sum_{i=1}^n {(N-N_i)}^2\cdot N_i$ in equation~\ref{eq:Results:CostGU}. The dominance of $t_{gu}$ is also highlighted in figure~\ref{img:TimeComparisons:CostRatio:NOSS} where the  \textit{cost ratio} follows a similar trend as the one observed in the global update in figure~\ref{img:TimeComparisons:GuCost}. The  \textit{cost ratio} can be maximised by choosing an appropriate $noss$ (close to where $t_{gu}$ is minimum) and increasing the ratio of prediction frequency, $pf$, to global update frequency, $guf$ (as seen in   equation~\ref{eq:Results:CostRatio}).
\begin{equation}
 t_{gu} \propto \sum\limits_{i=1}^n {(N-N_i)}^2\cdot N_i,\ \sum\limits_{i=1}^n {(N-N_i)}\cdot N_i^2,\ \sum\limits_{i=1}^n N_i^3
 \label{eq:Results:CostGU} 
\end{equation}

\begin{figure}[!htbp]
 \centering
 \includegraphics[width = .79\linewidth, keepaspectratio]{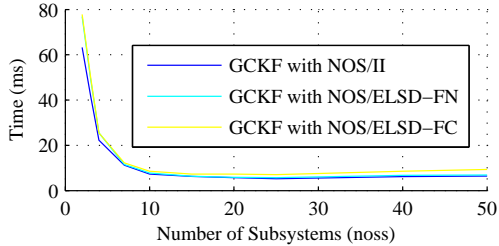}
 \caption{Comparison of $t_{gckf}$ of GCKF with NOS system archetype and different information exchange architectures against the number of subsystems ($noss$). The measured time is in the unit of milliseconds. The timing has been collected with the same parameters as figure~\ref{img:TimeComparisons:GuCost}.}
 \label{img:TimeComparisons:Cost:NOSS}
\end{figure}

The comparison of the processing cost of the GCKF ($t_{gckf}$) with various information exchange architectures under different number of subsystems ($noss$) is shown in figure~\ref{img:TimeComparisons:Cost:NOSS}. The increase in number of subsystems should, theoretically, decrease the amount of processing cost. However, due to  overhead in the implementation, the cost is seen to increase after reaching a minima (around $noss = 25$) in figure~\ref{img:TimeComparisons:Cost:NOSS}. The combination of the trends seen in figure~\ref{img:TimeComparisons:GuCost} and figure~\ref{img:TimeComparisons:Cost:NOSS}, together with the fixed ratio of $pf$ to $guf$, is responsible for the \textit{cost ratio} trend seen in figure~\ref{img:TimeComparisons:CostRatio:NOSS}. It is  also important to note that the cost of GCKF with II is lowest, followed by GCKF with ELSD-FN and ELSD-FC respectively. This order can also be seen in figure~\ref{img:TimeComparisons:CostRatio:NOSS}, where GCKF with II and GCKF with ELSD-FC has the maximum and minimum \textit{cost ratio} respectively.

The processing cost of the GCKF, $t_{gckf}$, will quadratically increase with the increase in the number of chosen frozen states, $noc$ in the ELSD-FC information exchange. This trend is because of the increase in dimensionality during the prediction step of the GCKF with ELSD-FC, from $N_k$ to $N_k + noc$. 

\begin{figure}[!htbp]
 \centering
 \includegraphics[width = \linewidth, keepaspectratio]{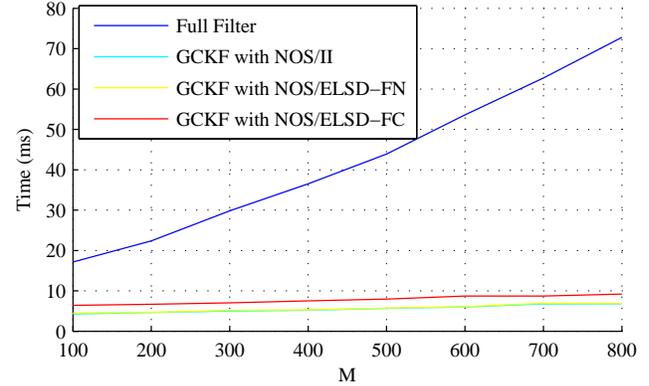}
 \caption{Comparison of $t_{ff}$ of full filter and $t_{gckf}$ of GCKF with NOS system archetype and different information exchange architectures against the overall observation dimensionality ($M$). The measured time is in the unit of milliseconds. The timing has been collected with following fixed parameters: $nos = 1000$; $noss = 20$; $nof = 4$.}
 \label{img:TimeComparisons:Cost:M}
\end{figure}

The processing cost for a standard estimation process (full filter) is proportional to $L\cdot N^2\cdot\sum_{i=1}^n M_i$, where $M_i$ is the dimensionality of the observation vector in subsystem $i$, and $L$ is the number of updates in the interval $\tatb$ \cite{JoseGCKF}. The application of the GCKF is justified for cases where the number of updates, $L$, or where the overall dimensionality of observations, $M$ ($=\sum_{i=1}^n M_i$), during the interval $\tatb$, are large. This can be seen by observing the growing disparity between $t_{ff}$ and $t_{gckf}$ in figure~\ref{img:TimeComparisons:Cost:M}; when the overall observation dimensionality, $M$, is increased. It is also worth mentioning that $L$ can be increased by increasing $pf$ (since $pf = uf$) which would, as previously mentioned, increase the ratio of $pf$ to $guf$, thereby increasing the  \textit{cost ratio} according to equation~\ref{eq:Results:CostRatio}.

\subsection{Non-linear Estimation With UKF Cores}\label{ssect:Results:UKF}
\begin{figure}[!htbp]
 \centering
 \begin{subfigure}{.33\linewidth}
  \centering
  \includegraphics[width = \linewidth, height = .758\linewidth]{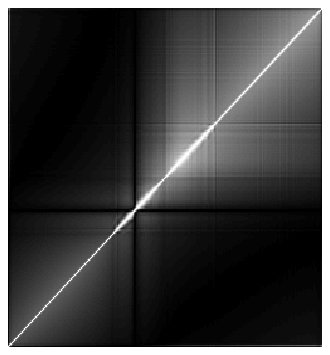}
  \caption{Full UKF}
  \label{subimg:ExploitingLocalBurgerSPV6:full}
 \end{subfigure}
 ~
 \begin{subfigure}{.5\linewidth}
  \centering
  \includegraphics[width = .66\linewidth, height = .5\linewidth]{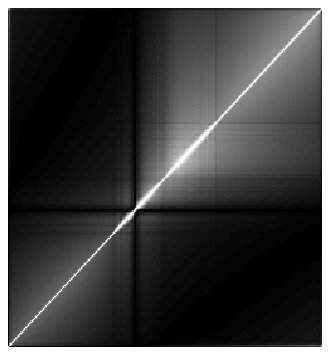}
  \caption{GCKF + NOS/ELSD-FC}
  \label{subimg:ExploitingLocalBurgerSPV6:NOS/ELSD-FC}
 \end{subfigure}
 \caption{Normalised covariance matrices: \subref{subimg:ExploitingLocalBurgerSPV6:full} full UKF and \subref{subimg:ExploitingLocalBurgerSPV6:NOS/ELSD-FC} GCKF with UKF core, ELSD-FC information exchange and NOS archetype. The results have been collected with the following parameters: $pf = 10^6$; $uf = 10^6$; $guf = 10^4$; $sf = 10^4$; $nos = 250$; $nof = 10$; $noss = 5$; $ol = [1,100]$; $\sigma^2_o = 25$; $\sigma^2_ s = 10 ^{-9}$.}
 \label{img:ExploitingLocalBurgerSPV6}
\end{figure}

\begin{figure}[!htbp]
 \centering
 \includegraphics[scale = 1.0]{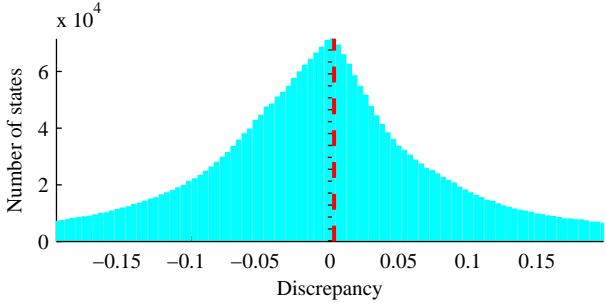}
 \caption{Comparison of discrepancy in expected value of GCKF with NOS/ELSD-FC similar to figure~\ref{img:ExploitingLocalHeatDB44}. The average discrepancy is recorded at $0.0030$.}
 \label{img:ExploitingLocalBurgerDB6}
\end{figure}

The Compressed estimation technique is general and allows treating certain class of non linear cases, which could be treated by a full UKF. The only requirement of the GCKF is that the resultant local belief, obtained from the individual estimation processes (by other variants of KF such as UKF), must be expressed via a Gaussian PDF; such that Gaussian virtual likelihood functions can be generated. The results from combining the NOS/ELSD-FC variant of GCKF with UKF cores, for solving one dimensional inviscid Burgers equation (equation~\ref{eq:Results:PDEs:Burgers}, discretised in $250$ cells and solved using Finite Volume numerical method) is shown in figures \ref{img:ExploitingLocalBurgerSPV6} and \ref{img:ExploitingLocalBurgerDB6}. Similar to the results in section~\ref{ssect:Results:ELSD}, the GCKF with NOS/ ELSD-FC was able to mimic the performance of the full UKF filter. Additionally, the average discrepancy of $0.1613$ is obtained when comparing the full UKF to EKF, highlighting the necessity of UKF in this problem.
\begin{equation}
 \frac{\partial U(x,t)}{\partial t} = -U(x,t)\frac{\partial U(x,t)}{\partial x}
 \label{eq:Results:PDEs:Burgers} 
\end{equation}
\section{Conclusion} \label{section:Conclusion}
This paper proposed new techniques to allow for compressed estimation of densely coupled estimation processes. The switching archetype and information exchange architectures (ELSD-FN and ELSD-FC) have been proposed with general theoretical formulations in either Bayesian or Gaussian perspective. The switching archetypes GCKF with ELSD information exchange architecture was concluded to have the best and acceptable performance, based on the simulations of different variants of GCKF in solving slow to intermediately fast 1D heat equation. The proposed techniques are general and when configured with appropriate Gaussian filter cores, such as UKF or CuKF, can be employed for compressed non-linear estimation. This has been briefly shown  for UKF in solving a 1D inviscid Burgers equation. The computational advantages of GCKF have also been highlighted and are maximally achieved when solving high dimensional and high frequency estimation problems.

\newcommand{\cond}[2]{(#1\vert #2)}
\newcommand{\Psubk}[2]{\bm{P}_{\bm{#1,(#2,#2)}}}
\appendix
 \section{Gaussian Virtual Update} \label{app:GVU}
 The augmented state vector for subsystem $k$ is modified from equation~\ref{eq:Review:stclone} to equation~\ref{eq:app:stclonenew}. The convention is consistent with \cite{JoseGCKF} and has been adopted for the convenience in defining the two likelihoods. 
\begin{equation}
 \setlength{\extrarowheight}{3.0pt}
 \Wktime[k]{t} = \transpose{[\transpose{\Wks{k}{1}}\timenota{t}, \transpose{\Wks{k}{2}}\timenota{t}]} = \transpose{[\Xkt{k}\timenota{\ta}, \Xkt{k}\timenota{t}]}
 \label{eq:app:stclonenew}
\end{equation}
The Gaussian PDF for the augmented system $k$ at time $t_a$ is initialised with the expected vector and the covariance matrix as follows,
\begin{equation}
\renewcommand*{\arraystretch}{1.0}
\begin{matrix}  
 \Wkh[k]\cond{\ta}{\ta} = 
 \left[
 \begin{array}{;{2pt/2pt}c;{2pt/2pt}}
  \hdashline \Xkh[k]\cond{\ta}{\ta}  \\
  \hdashline [?] \\ \hdashline
 \end{array} 
 \right];\ 
 \bm{P_k}\cond{\ta}{\ta} =
 \left[
 \begin{array}{;{2pt/2pt}c;{2pt/2pt}c;{2pt/2pt}}
  \hdashline \bm{A_0} & \bm{0} \\
  \hdashline \bm{0} & \bm{M} \\ \hdashline
 \end{array}
 \right] \\
 \bm{A_0} = \Psubk{k}{1}\cond{\ta}{\ta} \in \mathbb{R}^{N_k \times N_k} \\
  = \mathbb{E}\{(\Xktime[k]{\ta} - \Xkh[k]\cond{\ta}{\ta})\cdot\transpose{(\Xktime[k]{\ta} - \Xkh[k]\cond{\ta}{\ta})}\}\\
 \bm{M} = m\cdot\bm{I}_{N_k \times N_k},\ m \to \infty
 \end{matrix}
\end{equation} 
where, the symbol `?' means finite but irrelevant. 

The local estimates about $\Wktime[k]{\tb}$, of the individual subsystem $k$, obtained after the interval $\tatb$ are expressed by an expected value vector and a covariance matrix,
\begin{equation}
 \begin{matrix}
 \Wkh[k]\cond{\tb}{\tb} = 
 \left[
 \begin{array}{;{2pt/2pt}c;{2pt/2pt}}
  \hdashline \Wksh{k}{1}\cond{\tb}{\tb} \\
  \hdashline \Wksh{k}{2}\cond{\tb}{\tb} \\ \hdashline
 \end{array}
 \right];\ 
 \bm{P_k}\cond{\tb}{\tb} = 
 \left[ 
 \begin{array}{;{2pt/2pt}c;{2pt/2pt}c;{2pt/2pt}}
  \hdashline \bm{A} & \transpose{\bm{C}} \\
  \hdashline \bm{C} & \bm{B} \\ \hdashline
 \end{array}
 \right]\\
 \bm{A} = \Psubk{k}{1}\cond{\tb}{\tb};\  
 \bm{B} = \Psubk{k}{2}\cond{\tb}{\tb};\ \bm{A},\bm{B} \in \mathbb{R}^{N_k \times N_k} 
 \end{matrix}
\end{equation}

The individual global update is implemented through a sequence of two virtual update steps. The constrained virtual update is done based on the change in the component $\Wks{k}{1}$ (i.e.\ the stochastic clone of $\Xktime[k]{\ta}$) and is implemented by considering,
\begin{equation}
 \begin{aligned}
  \bigtriangleup\Xkh[a] &= \Wksh{k}{1}\cond{\tb}{\tb} - \Wksh{k}{1}\cond{\tb}{\ta}  \\
  \bigtriangleup\bm{ P_{a,a}} &= \Psubk{k}{1}\cond{\tb}{\ta} - \Psubk{k}{1}\cond{\tb}{\tb} = \bm{A_0} - \bm{A}
 \end{aligned}
\end{equation}

The second virtual update, also known as the uninformative virtual update, has the mission of replacing the random variable $\Xktime[k]{\ta}$ by the random variable which describes $\Xktime[k]{\tb}$. This update does not provide any information to the rest of the system, as it only affects the marginal PDF about the random variable $\Xktime[k]{\tb}$ and its statistical dependency with the rest of the random variables. This second update is applied through a virtual prediction for the subset of states $\Xk[k]$ by the process model,
\begin{equation}
 \begin{aligned}
  \Xktime[k]{\ta} &\to \Xktime[k]{\tb} \\
  \Xktime[k]{\tb} &= G(\Xktime[k]{\ta}) + \bm{\xi}\timenota{t} \\ &= 
  \bm{C}\cdot\bm{A}^{-1}\cdot(\Xktime[k]{\ta} - \Wksh{k}{1}\cond{\tb}{\tb}) \\ &\ + \Wksh{k}{2}\cond{\tb}{\tb} + \bm{\xi}\timenota{t}
 \end{aligned}
 \label{eq:app:VirtualPrediction}
\end{equation}
where $\bm{\xi}\timenota{t}$ is considered to be zero mean GWN, whose covariance is $\bm{Q_{\xi}} = \bm{B} - \bm{C}\cdot\bm{A}^{-1}\cdot\transpose{\bm{C}} \geq \bm{0}$

These two steps produce the exact overall results in comparison with the standard virtual update, but at the cost of avoiding the temporary expansion of full state vector and its subsequent marginalisation step which were present in the general Bayesian case. 

The information regarding Degenerated Gaussian Observation(DGO) and virtual observation when $\bigtriangleup\bm{ P_{a,a}}$ is singular can be found in \cite{JoseGCKF}.

\section{Decorrelation Method}\label{app:Decorrelation}
In order to conservatively decorrelate a subset of estimates, a proper increase in the value of diagonal sub-matrices of a covariance matrix is sufficient. The decorrelation method presented here is a simple application of \cite{JoseMem} for multiple subsystems, where the multiplicative decorrelation constant for the individual subsystem is formulated. 

Let the number of interested subsystems be $V$, the method in \cite{JoseMem} can be applied by dividing $V$ subsystems into two groups for decorrelating. One group would contain one subsystem while the other group would contain $V-1$ subsystems. This step is then repeated in the divided groups until all individual $V$ subsystems are decorrelated. The decorrelation technique for two group of states in \cite{JoseMem} is shown in equation~\ref{eq:app:GenDecor}, where $\bm{P}$ represents the original covariance matrix and $\bm{P}^*$ represents the decorrelated covariance matrix with decorrelation coefficient $c$. If $c$ is initially chosen to be $V-1$ and is iteratively decreased by one in the subsequent decorrelation steps (i.e.\ $V-2, V-3, \ldots, 1$), the final multiplicative constant to achieve the diagonal blocks in the covariance matrix can be shown to be simply $V$. The semi-positive definite decorrelated covariance matrix can then be obtained by using equation~\ref{eq:app:DecorrelatedCo}.
\begin{equation}
\begin{aligned}
 \bm{P} &\leq \bm{P}^*,\ \forall c>0 \\
 \left[
 \begin{array}{;{2pt/2pt}c;{2pt/2pt}c;{2pt/2pt}}
 	\hdashline \bm{P}_{11} & \bm{P}_{12} \\ \hdashline
 	\bm{P}_{21} & \bm{P}_{22} \\ \hdashline
 \end{array}
 \right] &\leq \left[
 \begin{array}{;{2pt/2pt}c;{2pt/2pt}c;{2pt/2pt}}
 	\hdashline \bm{P}_{11}\cdot (1+c) & \bm{0} \\ \hdashline
 	\bm{0} & \bm{P}_{22}\cdot (1+ \frac{1}{c}) \\ \hdashline
 \end{array}
 \right]
 \end{aligned}	   
 \label{eq:app:GenDecor}
\end{equation}
\begin{equation}
 \bm{P}^* = \bigoplus\limits_{i=1}^V V\cdot \bm{P}_{[i]}
 \label{eq:app:DecorrelatedCo}
\end{equation}

\section{General Statistical Functional Relationship}\label{app:GenFunc}
For a certain family of PDFs (which includes the Gaussian cases), if a set of random variables, $\bm{Z} = \transpose{[\transpose{\bm{Z_a}},\transpose{\bm{Z_b}}, \transpose{\bm{Z_c}}]}$, that is described by a joint PDF (which is a memeber of that family), $p(\bm{Z})$, the subset of states $\bm{Z_c}$ can be statistically represented as a function of the states $\bm{Z_a}$ via the expression $\bm{Z_c} = G_a(\bm{Z_a}) + \bm{\zeta}$. The uncertainty component $\bm{\zeta}$, although independent of the states $\bm{Z_a}$, may have statistical dependency with the states $\bm{Z_b}$, i.e.\ $\bm{\zeta} = G_b(\bm{Z_b}) + \bm{\zeta_b}$, where the noise component $\bm{\zeta_b}$ is independent of both the states $\bm{Z_a}$ and $\bm{Z_b}$. In cases where there is strong statistical dependency between the states $\bm{Z_a}$ and $\bm{Z_b}$, the uncertainty due to the unmodeled component $G_b(\bm{Z_b})$ will be marginal, i.e.\ $\bm{\zeta} \approx \bm{\zeta_b}$. Although in this paper $\bm{\zeta}$ is assumed to be GWN, the method of covariance intersection, proposed in \cite{julier1997non}, should be applied due to the presence of unknown correlation between $\bm{\zeta}$ and $\bm{Z_b}$.

In relation to the information exchange architecture ELSD (Exploiting Local Statistical Dependency), the functional relationship $G_a$ and the covariance of the noise component $\bm{\zeta}$ is obtained through statistical linear regression, i.e.\ $\bm{Z_c} = \bm{\alpha}\cdot\bm{Z_a} + \bm{b} + \bm{\zeta}$. This is shown in equation~\ref{eq:IEA:TransformationFrozen}, where $\bm{Z_c}$ represents the states in subsystem $m_d$ that are required in the sub-process $k$ (i.e.\ $\Xnds{k}{d}\timenota{t}$) while $\bm{Z_a}$ represents the states that are common in subsystems $m_d$ and $k$, such that the local statistical dependency can be exploited. One example is a particular case of ELSD, where $\bm{Z_a}$ is the time-invariant version of  $\bm{Z_c}$. This version of ELSD information exchange (referred to as ELSD-FN) exploits the statistical dependency between $\Xnds{k}{d}\timenota{t}$ and $\Xnds{k}{d}\timenota{\ta}$.
 
\section{1D Heat PDE}\label{app:PDEs}
Consider the temperature $U(x,t)$ in a rod where the temperature is governed by the heat equation given in equation~\ref{eq:app:heat} with Dirichlet boundary condition. The variable $x$ denotes spatial dimension  while $t$ denotes the temporal dimension of temperature.
\begin{equation}
 \frac{\partial U(x,t)}{\partial t} = \beta \frac{\partial^2 U(x,t)}{\partial x^2}
 \label{eq:app:heat} 
\end{equation}
The variable $\beta$ in equation~\ref{eq:app:heat} is defined by the properties of the rod: thermal conductivity ($k$), density ($\rho$), and specific heat ($C_p$).
\begin{equation}
 \beta = {k}\times{(\rho \times C_p)}^{-1}
 \label{eq:app:beta}
\end{equation}

The two ends of the rod are kept at $0 \SI{}{\celsius}$ and $400\SI{}{\celsius}$ respectively and the initial temperature of the rod is assumed to be $23 \SI{}{\celsius}$. The sides of the rod are assumed to be insulated  with constant, uniform and temperature independent properties: specific heat, density, thermal conductivity and cross sectional area.
\begin{figure}[!htbp]
\centering
\begin{tikzpicture}
 \edef\l{2}
 \edef\h{0.5}
 \edef\t{2.0}
 \edef\th{0.75}
 \edef\sf{1.5}
 \edef\se{0}
 \edef\le{0.2}
 \node (A) at (\sf, \se) {$U(0,t) = \SI{0}{\celsius}$};
 \node (B) at (\sf + \t,\se) {};
 \node (C) at (\sf + 2*\t + \l,\se) {$U(l,t) = \SI{400}{\celsius}$}; 
 \node (D) at (\sf + \t + \l, \se) {};
 \node (E) at (\sf + \t + \l/2, \se - \h/2 - \th) {$U(x,0) = \SI{23}{\celsius}$};
 \node (F) at (\sf + \t + \l/2, \se - \h/2) {};
 \node (G) at (\sf + \t, \se - \h/2 - \le) {$x = 0$};
 \node (H) at (\sf + \t + \l, \se - \h/2 - \le) {$x = l$};
 \draw (\sf + \t,\se - \h/2) rectangle (\sf + \t + \l,\se + \h/2);
 \draw [->] (A) -- (B);
 \draw [->] (C) -- (D);
 \draw [->] (E) -- (F);
\end{tikzpicture}
\caption{Representation of the initial conditions of the rod}
\label{img:Rod}
\end{figure}

The discrete time $t$ and spatial variable $x$ are given by,
\begin{equation}
\begin{aligned}
 0 \leq x \leq l;\ \Delta_x = \frac{l}{n+1};\  x_j = j\Delta_x;\ 0 \leq j \leq n+1 \\ 
 0 \leq t \leq T;\ \Delta_t = \frac{T}{m};\ t_k = k\Delta_t;\ 0 \leq k \leq  m 
\end{aligned}
\label{eq:app:DiscreteVars}
\end{equation}
where, $n$ and $m$ represent the total number of discretised states in the spatial dimension and the temporal dimension respectively. If $U_j^k \vcentcolon= U(x_j,t_k)$, using forward-time-centred-space (FTCS) scheme, the discretised version of equation~\ref{eq:app:heat} can be written as,
\begin{equation}
\begin{aligned}
U_j^{k+1} &= s(U_{j+1}^k + U_{j-1}^k) + (1-2s)U_j^k \\
s &= {(\beta \Delta_t)}/{(\Delta_x^2)}
\end{aligned}
\label{eq:app:FTCS}
\end{equation}

The variable $s$ in equation~\ref{eq:app:FTCS} quantifies the rate of change of temperature in the discretised states in one time step. This variable will be referred from henceforth as an indicator for the dynamic of the system with the stability constraint, $s \leq 0.5$. 

The sensor readings that were used as an observation for the filter are simulated by adding white Gaussian noise to the solution of equation~\ref{eq:app:heat}, which was obtained numerically from the method of line.

\bibliographystyle{elsarticle-num}
\bibliography{references}

\end{document}